\documentclass[aps,prd,twocolumn,showpacs,byrevtex]{revtex4-1}
\usepackage{times, amsmath, graphicx, color, overpic, enumitem}
\usepackage{multirow, lineno, verbatim, setspace}
\usepackage{ulem}
\usepackage{xspace, bm}
\usepackage[pdfstartview=FitH, colorlinks, urlcolor=blue,
citecolor=blue,linkcolor=blue]{hyperref}
\usepackage[T1]{fontenc}

\graphicspath{{./figure/}}     
\DeclareGraphicsExtensions{.pdf,.jpeg,.png}

\newcommand{\jpsi}{J/\psi}
\newcommand{\psip}{\psi(3686)}
\newcommand{\p}{p}
\newcommand{\pb}{\bar{p}}
\newcommand{\pbar}{\bar{p}}
\newcommand{\etap}{\eta^{\prime}}
\newcommand{\csq}{c^{2}}
\newcommand{\pip}{\pi^{+}}
\newcommand{\pim}{\pi^{-}}

\newcommand{\br}{\mathcal{B}}
\newcommand{\elep}{e^{+}}
\newcommand{\elem}{e^{-}}
\newcommand{\chisq}{\chi^{2}}

\newcommand{\mm}{\mathcal{M}}
\newcommand{\mf}{\Omega}
\newcommand{\bpsi}{\ensuremath{\bm{\psi}}\xspace}

\begin{document}


\title{\boldmath{Observation of $\psip\to\p\pb\etap$ and improved measurement of $\jpsi\to\p\pb\etap$}}

\author{
\begin{small}
\begin{center}
M.~Ablikim$^{1}$, M.~N.~Achasov$^{10,d}$, S. ~Ahmed$^{15}$, M.~Albrecht$^{4}$, M.~Alekseev$^{55A,55C}$, A.~Amoroso$^{55A,55C}$, F.~F.~An$^{1}$, Q.~An$^{52,42}$, J.~Z.~Bai$^{1}$, Y.~Bai$^{41}$
% ...
\\
\vspace{0.2cm}
(BESIII Collaboration)\\
\vspace{0.2cm} {\it
$^{1}$ Institute of High Energy Physics, Beijing 100049, People's Republic of China\\
% ...
$^{4}$ Bochum Ruhr-University, D-44780 Bochum, Germany\\
% ...
$^{10}$ G.I. Budker Institute of Nuclear Physics SB RAS (BINP), Novosibirsk 630090, Russia\\
% ...
$^{15}$ Helmholtz Institute Mainz, Johann-Joachim-Becher-Weg 45, D-55099 Mainz, Germany\\
% ...
$^{41}$ Southeast University, Nanjing 211100, People's Republic of China\\
$^{42}$ State Key Laboratory of Particle Detection and Electronics, Beijing 100049, Hefei 230026, People's Republic of China\\
% ...
$^{52}$ University of Science and Technology of China, Hefei 230026, People's Republic of China\\
% ...
$^{55}$ (A)University of Turin, I-10125, Turin, Italy; (B)University of Eastern Piedmont, I-15121, Alessandria, Italy; (C)INFN, I-10125, Turin, Italy\\
% ...
$^{d}$ Also at the Novosibirsk State University, Novosibirsk, 630090, Russia\\
% ...
}\end{center}
\vspace{0.4cm}
\end{small}
}
\noaffiliation{}

\date{\today}

\begin{abstract}
  We observe the process $\psip\to\p\pb\etap$ for the first time, with a statistical significance higher than 10$\sigma$, and measure the branching fraction of $\jpsi\to\p\pb\etap$ with an improved accuracy compared to earlier studies. The measurements are based on $4.48 \times 10^8$ $\psip$ and $1.31 \times 10^{9}$ $\jpsi$ events collected by
  the BESIII detector operating at the BEPCII. The branching fractions are determined to be $\br(\psip\to\p\pb\etap) = (1.10\pm0.10\pm0.08)\times10^{-5}$ and
  $\br(\jpsi\to\p\pb\etap)=(1.26\pm0.02\pm 0.07)\times10^{-4}$, where the first uncertainties are statistical and the second ones systematic. Additionally, the $\eta-\eta^{\prime}$ mixing angle is determined to be $-24^{\circ} \pm 11^{\circ}$ based on $\psip \to \p \pb \etap$, and $-24^{\circ} \pm 9^{\circ}$ based on $\jpsi \to \p \pb \etap$, respectively.
\end{abstract}


\maketitle

\section{INTRODUCTION}

\par Quantum Chromodynamics (QCD), the theory describing the strong interaction, has been tested thoroughly at high energy. However, in the medium energy region, theoretical calculations based on first principles are still unreliable since the non-perturbative contribution is significant and calculations have to rely on models. Experimental measurements in this energy region are helpful to validate or falsify models, constrain parameters, and inspire new calculations. Charmonium states are on the boundary between perturbative and non-perturbative regimes in QCD, therefore, their decays, especially the hadronic decays, provide ideal inputs to study the QCD. The availability of very large samples of vector charmonia, produced via electron-positron annihilation, such as $\jpsi$ and $\psip$, makes experimental studies of rare processes and decay channels with complicated intermediate structures possible.

\par Among these hadronic decays, scenarios of {\boldmath $\psi$} (in the following, $\bpsi$ denotes either $J/\psi$ or $\psi(3686)$) decaying into baryon pairs have been understood via $c \bar{c}$ annihilation into three gluons or a virtual photon~\cite{psi_to_BBbar}. But its natural extension, the three-body decays, $\bpsi \to\p\pb P$, where $P$ represents a pseudoscalar meson such as $\pi^0$, $\eta$, or $\eta'$, still need more studies since intermediate states contribute significantly here. Specific models based on nucleon and $N^*$ pole diagrams have been proposed to deal with these problems~\cite{soft_pion_theorem,nucleon_pole,nucleon_resonance}. However, recent studies have focused on the final states $\p\pb \pi^0$ and $\p \pb \eta$, and not so much on $\p\pb \eta'$, partially due to the limited experimental measurements.

\par The study of the process $\bpsi\to\p\pb \eta'$, as well as the branching fraction of $\bpsi\to\p\pb\eta$, can also be used to determine the $\eta-\etap$ mixing angle $\theta_{\eta-\etap}$. The $\eta-\etap$ mixing angle, which was proposed in quark model SU(3) flavor symmetry~\cite{soft_pion_theorem}, is expected to be $-(10^{\circ} \sim 17^{\circ})$ based on a QCD inspired calculation~\cite{soft_pion_theorem} or $-(13^{\circ}\sim 16^{\circ})\pm6^{\circ}$ based on the quark-line rule (QLR)~\cite{quark-line-rule}.

\par In addition, using the process $\bpsi \to \p\pb \eta'$, we are able to test the ``12\% rule''. The ratio $Q$ of the branching fractions of $J/\psi$ and $\psi(3686)$ can be written in terms of their total and leptonic widths under the assumption that the charmonium systems are non-relativistic and decay to hadrons predominantly via point-like annihilation into three gluons~\cite{12_rule_01, Augustin:1974xw}: $Q = \frac{\br(\psip\to ggg)}{\br(\jpsi\to ggg)} = \frac{\Gamma(\psip\to\elep\elem) \cdot \Gamma(\jpsi)}{\Gamma(\jpsi\to\elep\elem) \cdot \Gamma(\psip)} = (12.2\pm 2.4)\%$~\cite{rhopi_puzzle_mrkii}. This relation was extended to exclusive processes later, ignoring other factors associated with each exclusive mode such as multiplicity and phase space factors. Although the ``12\% rule'' has been confirmed experimentally for many decay modes, severe violation has been found in several channels~\cite{rhopi_puzzle_mrkii}. Many theoretical explanations~\cite{rho_pi_rev} have been proposed to explain the violation of the ``12\% rule'', but none is satisfactory.

\par The baryonic three-body decay $\jpsi\to\p\pb \eta'$ was first observed by MARKI with branching fraction $(1.8 \pm 0.6) \times 10^{-3}$ in 1978~\cite{jpsi_ppbetap_mrki} and confirmed by MARKII with branching fraction $(0.68 \pm 0.23 \pm 0.17) \times 10^{-3}$ in 1984~\cite{jpsi_ppbetap_mrkii}. Later, using $5.80 \times 10^7$ $\jpsi$ events, BESII performed a further measurement of the branching fraction of $\jpsi\to\p\pb \eta'$ with $(2.00 \pm 0.23 \pm 0.28) \times 10^{-4}$~\cite{jpsi_ppbetap_besii}. The process $\psip \to \p \pb \etap$ has not been observed so far.

\par At present, two large samples of $4.48 \times 10^8$ $\psip$ events~\cite{psip_num} and $1.31 \times 10^{9}$ $\jpsi$ events~\cite{jpsi_num} have been collected with the BESIII detector. Using these large data sets, we present the first observation of $\psip\to\p\pb\etap$ and an improved measurement of the branching fraction of $\jpsi\to\p\pb\etap$. In this analysis, the $\etap$ candidates are reconstructed via their decays $\etap\to\gamma\pip\pim$ and $\etap\to\pip\pim\eta, \eta \to \gamma \gamma$.

\section{BESIII DETECTOR AND DATA SAMPLES}

\par The BESIII detector, described in detail in Ref.~\cite{detail_bepcii}, has a geometrical acceptance of 93\% of 4$\pi$ solid angle.
It can be subdivided into four main sub-detectors. A helium-based multi-layer drift chamber (MDC) determines the momentum of charged particles,
traveling through a 1~T (for $\jpsi$ sample 0.9~T in 2012) magnetic field, with a resolution 0.5\% at 1~GeV/$c$, as well as the ionization energy loss (d$E$/d$x$) with a resolution better than 6.0\% for electrons from Bhabha scattering. A time-of-flight system (TOF) made of plastic scintillators with a time resolution of 80 ps (110 ps) in the barrel (end caps) is used for particle identification (PID). An electromagnetic calorimeter (EMC) consisting of 6240 CsI(Tl) crystals measures the energies of photons with a resolution of 2.5\% (5.0\%) in the barrel (end caps) at 1~GeV, and their positions with a resolution of 6 mm (9 mm) in the barrel (end caps). A muon counter (MUC) based on resistive plate chambers with 2~cm position resolution provides information for muon identification.

\par A {\sc{geant4}}-based~\cite{geant4_01, geant4_02} detector simulation package is used to model the detector response. Inclusive Monte Carlo (MC) samples,
including $5.06 \times 10^8$ $\psip$ and $1.23 \times 10^9$ $\jpsi$ events, are used for background studies. The production of the $\psip$ and $\jpsi$ resonances is
simulated using the event generator {\sc{kkmc}}~\cite{kkmc_01, kkmc_02}, and their decays are simulated by
{\sc{evtgen}}~\cite{evtgen} for those with a known branching fraction obtained from the Particle Data Group (PDG)~\cite{pdg_2014} and by the {\sc{lundcharm}} model~\cite{lundcharm} for unmeasured ones. Signal MC samples are generated to determine the detection efficiency and to optimize selection criteria. The $\psip\to\p\pb\etap$, $\jpsi\to\p\pb\etap$, $\etap\to\eta\pip\pim$ and $\eta\to\gamma\gamma$ are generated according to phase space (PHSP) distributions, while $\etap\to\gamma\pip\pim$ according to the model-dependent amplitudes determined in Ref.~\cite{diy_test3}. Data samples taken at center-of-mass energies $\sqrt{s}=3.080$ and $3.650$~GeV with integrated luminosities of $31$~pb$^{-1}$
and $44$~pb$^{-1}$, respectively, are used to estimate continuum backgrounds.

\section{EVENT SELECTIONS AND BACKGROUND ANALYSIS}

\par Charged tracks are reconstructed from hits in the MDC. For each track, the polar angle must satisfy $|\cos\theta|<0.93$ and the point of closest approach to the interaction point must be within $\pm1$~cm in the plane perpendicular to the beam and $\pm10$ cm along the beam direction. The TOF and d$E$/d$x$ information is combined to calculate PID likelihoods for the pion and proton hypotheses, and the PID hypothesis with the largest likelihood is assigned to the track.

\par Photons are reconstructed from isolated electromagnetic showers in the EMC. The angle between the directions of any cluster and its nearest charged
track must be larger than 10 or 30 degrees to pion or (anti-)proton tracks, respectively. The efficiency and energy resolution are improved by
including the energy deposited in nearby TOF counters. A photon candidate must deposit at least 25~MeV (50~MeV) in the barrel (end caps) region,
corresponding to an angular coverage of $|\cos\theta|<0.80$ ($0.86<|\cos\theta|<0.92$). The timing information obtained from the EMC is required to
be $0\leq t_\text{EMC}\leq 700$~ns to suppress electronic noise and beam backgrounds unrelated to the event.

\par Signal candidates must have four charged tracks identified as $\p$, $\pb$, $\pip$ and $\pim$, as well as at least one (two) photon(s) for the $\etap\to\gamma\pip\pim\ (\etap\to\eta\pip\pim)$ mode. The events with $920 < M_{\gamma\pip\pim(\eta\pip\pim)} < 1000$~MeV/$\csq$ are accepted for further analysis, where $M_{\gamma\pip\pim(\eta\pip\pim)}$ is the invariant mass of $\gamma \pip \pim$ or $\eta \pip \pim$, respectively. To improve the resolution and to suppress backgrounds, a kinematic fit to all final state particles with a constraint on the initial $e^+ e^-$ four-momentum is performed. In addition, for the $\etap\to\pip\pim\eta$ mode, the invariant mass of the two photons is constrained to the nominal mass of $\eta$. The $\chisq$ of the kinematic fit for each decay mode is required to be less than an optimized value obtained by maximizing the figure of merit $S/\sqrt{S+B}$, where $S$ is the number of signal events from a signal MC sample normalized to the preliminary measurements and $B$ is the number of background events in the $\etap$ signal region obtained from inclusive MC samples. When there are more photon candidates than required in an event, we loop over all possible combinations and keep the one with the smallest kinematic fit $\chisq$. After the kinematic fit, the $\etap$ signal region is defined as $948.2<M_{\gamma\pip\pim(\eta\pip\pim)}<967.4$~MeV/$\csq$, while the sideband regions are defined as $932.4 < M_{\gamma\pip\pim(\eta\pip\pim)} < 942.0$~MeV/$\csq$ and $974.0<M_{\gamma\pip\pim(\eta\pip\pim)}<983.6$~MeV/$\csq$.

To remove background events, we apply the following requirements:
\begin{itemize}[leftmargin=*]
  \addtolength{\itemsep}{-5pt}
  \item For $\psip \to p \pb \etap $ with $\etap \to \gamma \pip \pim$, we veto $\psip\to\gamma\chi_{cJ}, \chi_{cJ} \to p \pb \pip \pim$ ($J$ = 0, 1, 2) decays by requiring $|M_{\gamma}^{\rm rec}-m_{\chi_{c0}}|>40$ MeV/$c^{2}$, $|M_{\gamma}^{\rm rec}-m_{\chi_{c1}}|>14$ MeV/$c^{2}$ and $|M_{\gamma}^{\rm rec}-m_{\chi_{c2}}|>15$ MeV/$c^{2}$, and veto $\psip\to\pip\pim\jpsi, \jpsi \to \gamma p \pb $ decays by requiring $|M^{{\rm{rec}}}_{\pip\pim}-m_{\jpsi}|>8$ MeV/$\csq$.
  \item For $\psip \to p \pb \etap$ with $\etap \to \eta \pip \pim $, we veto $\psip\to\eta\jpsi, \eta \to \gamma \gamma,  \jpsi \to p \pbar \pip \pim$ decays by requiring $|M^{{\rm{rec}}}_{\gamma\gamma}-m_{\jpsi}|>8$ MeV/$\csq$, and veto $\psip\to\pip\pim\jpsi, \jpsi \to \eta p \pb $ decays by requiring $|M^{{\rm{rec}}}_{\pip\pim}-m_{\jpsi}|>8$ MeV/$\csq$.
\end{itemize}
Here, $M_{\gamma}^{\rm rec}$, $M^{{\rm{rec}}}_{\pip\pim}$, and $M^{{\rm{rec}}}_{\gamma\gamma}$ are the recoil mass of $\gamma$, $\pip\pim$, and $\gamma\gamma$, while $m_{\chi_{cJ}}$ and $m_{\jpsi}$ are the nominal $\chi_{cJ}$ and $\jpsi$ masses~\cite{pdg_2014}, respectively. The mass window for each requirement is determined based on the exclusive MC simulation.

\par The backgrounds from $\psip$ and $\jpsi$ decays are studied with inclusive MC samples. For $\psip \to p \pb \etap $ with $\etap \to \gamma \pip \pim$, even after the $\chi_{cJ}$ mass window requirements, the main remaining background is the decay $\psip\to\gamma\chi_{cJ}$ with $\chi_{cJ}\to\p\pb\pip\pim$, which has the same final state as the signal process. A study of MC simulated events shows that the $M_{\gamma\pip\pim}$ distribution from $\psip\to\gamma\chi_{cJ}\to\gamma\p\pb\pip\pim$ is smooth.  Therefore, its contribution can be easily determined in a fit. For the other three decay modes, $\psip \to p \pb \etap$ with $\etap \to \eta \pip \pim$, $\jpsi \to p \pb \etap$ with $\etap \to \gamma \pip \pim$ and $\etap \to \eta \pip \pim$, there are no dominant background processes, but many decay channels with a small contribution each. The backgrounds from the continuum process $\elep\elem\to q \bar{q}$ are studied with data samples taken at $\sqrt{s}=3.080$ and $3.650$~GeV.  The background level is found to be very low, and the background events do not peak in the signal region.

\begin{figure}[hbtp]
  \centering
  \begin{overpic}[width=0.23\textwidth]{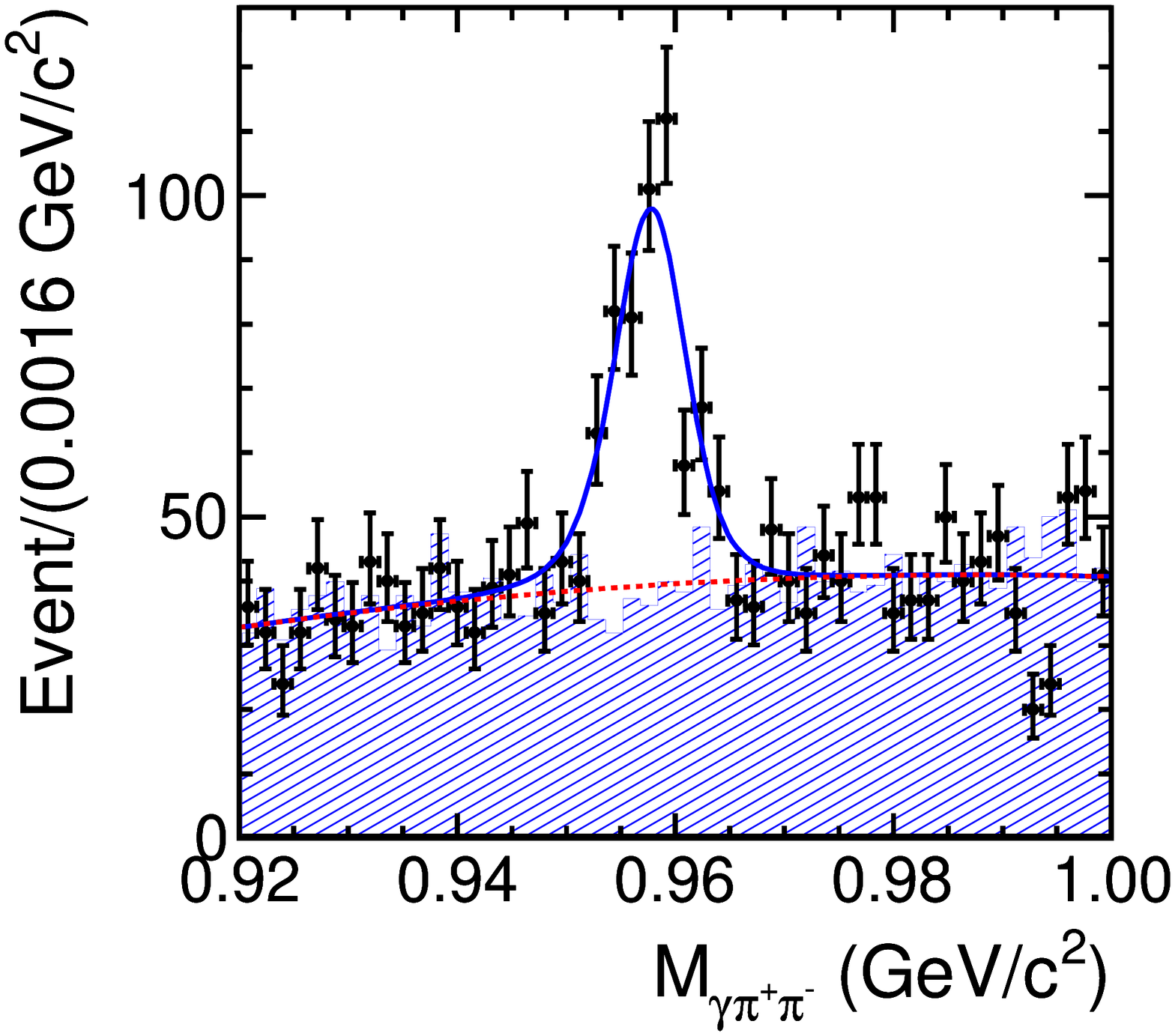}
    \put(80, 80){(a)}
  \end{overpic}
  \begin{overpic}[width=0.23\textwidth]{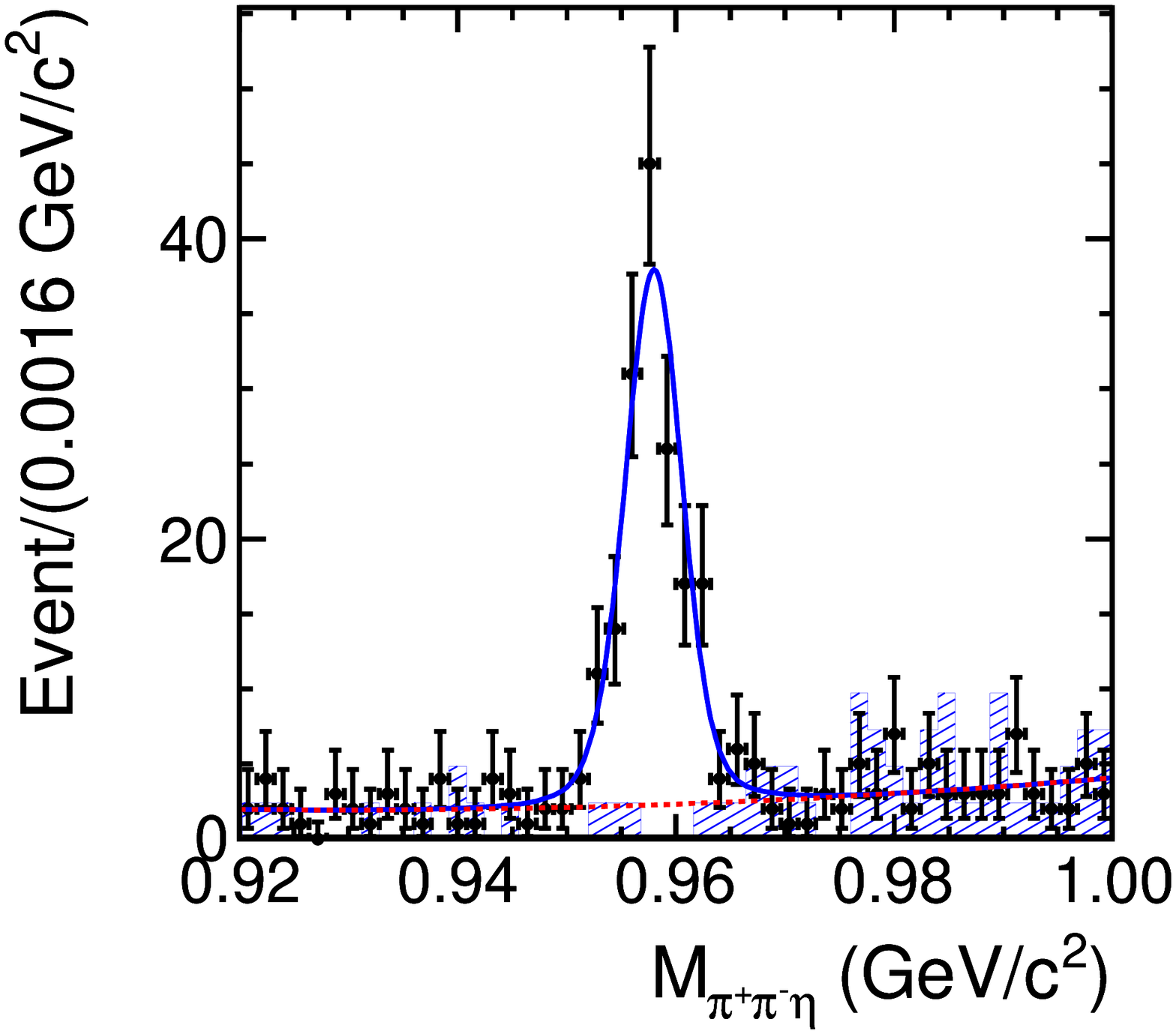}
    \put(80, 80){(b)}
  \end{overpic}
  \begin{overpic}[width=0.23\textwidth]{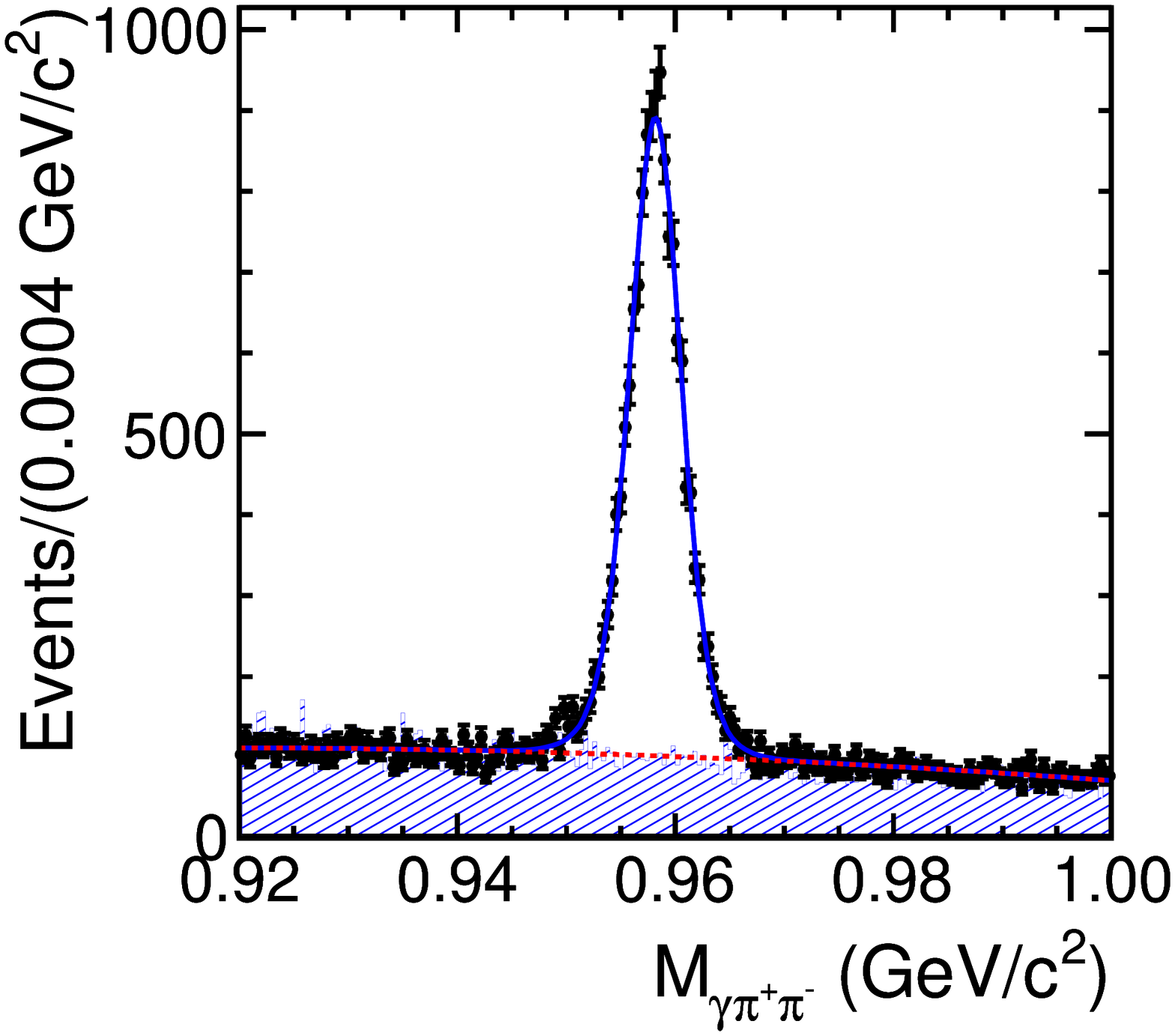}
    \put(80, 80){(c)}
  \end{overpic}
  \begin{overpic}[width=0.23\textwidth]{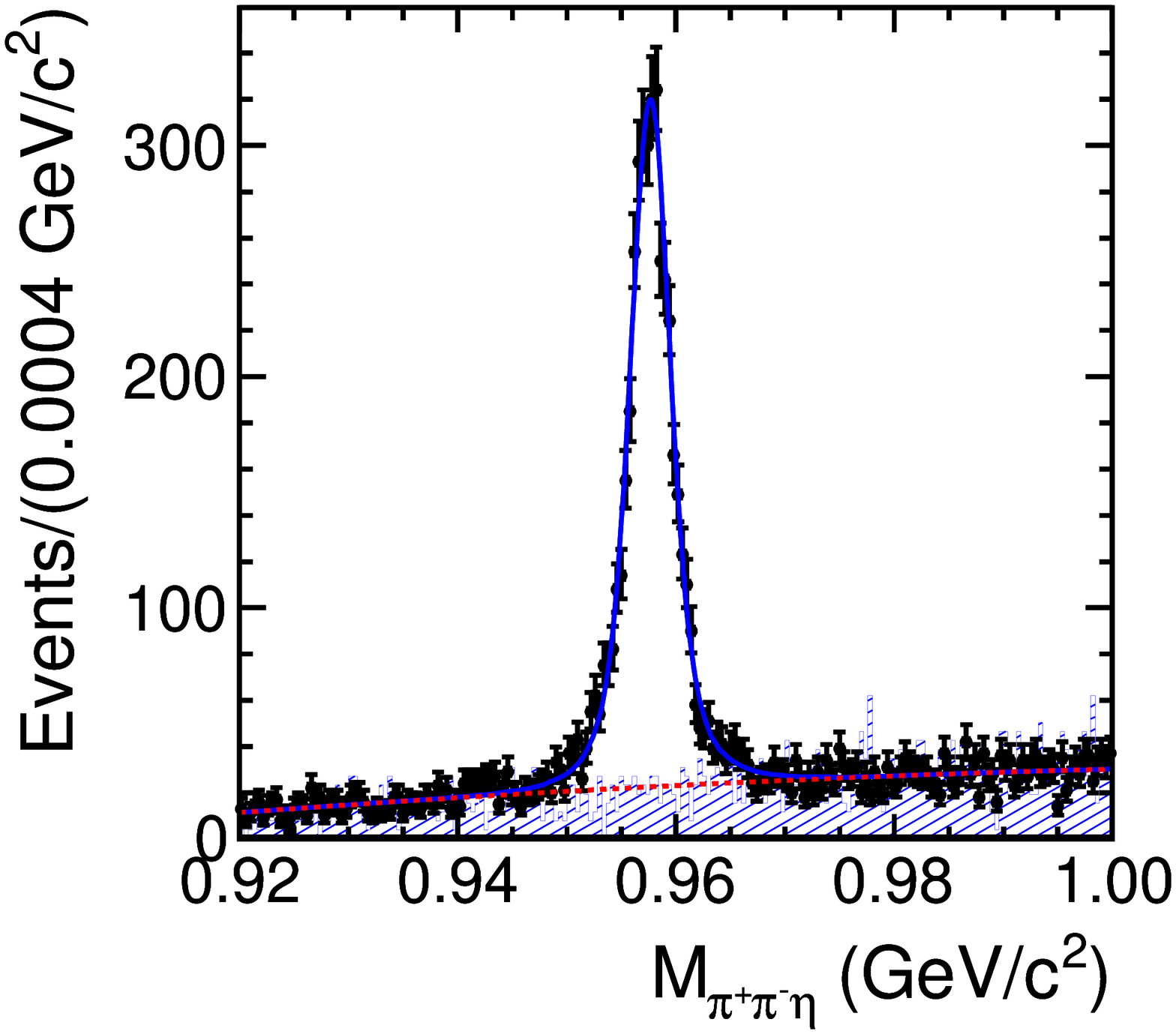}
    \put(80, 80){(d)}
  \end{overpic}\\
  \caption{Invariant mass spectra of the $\etap$ candidates in $\psip\to\p\pb\etap$ with $\etap\to\gamma\pip\pim$ (a), $\psip\to\p\pb\etap$ with $\etap\to\eta\pip\pim$ (b), $\jpsi\to\p\pb\etap$ with $\etap\to\gamma\pip\pim$ (c), and $\jpsi\to\p\pb\etap$ with $\etap\to\eta\pip\pim$ (d). The dots with error bars are data, the  shaded histograms are the backgrounds from inclusive MC samples, the blue solid curves are the fit results, and the red dashed curves are the backgrounds from fit.}
  \label{fig-fit-invm-etap}
\end{figure}

\begin{figure}[hbtp]
  \centering
   \begin{overpic}[width=0.23\textwidth]{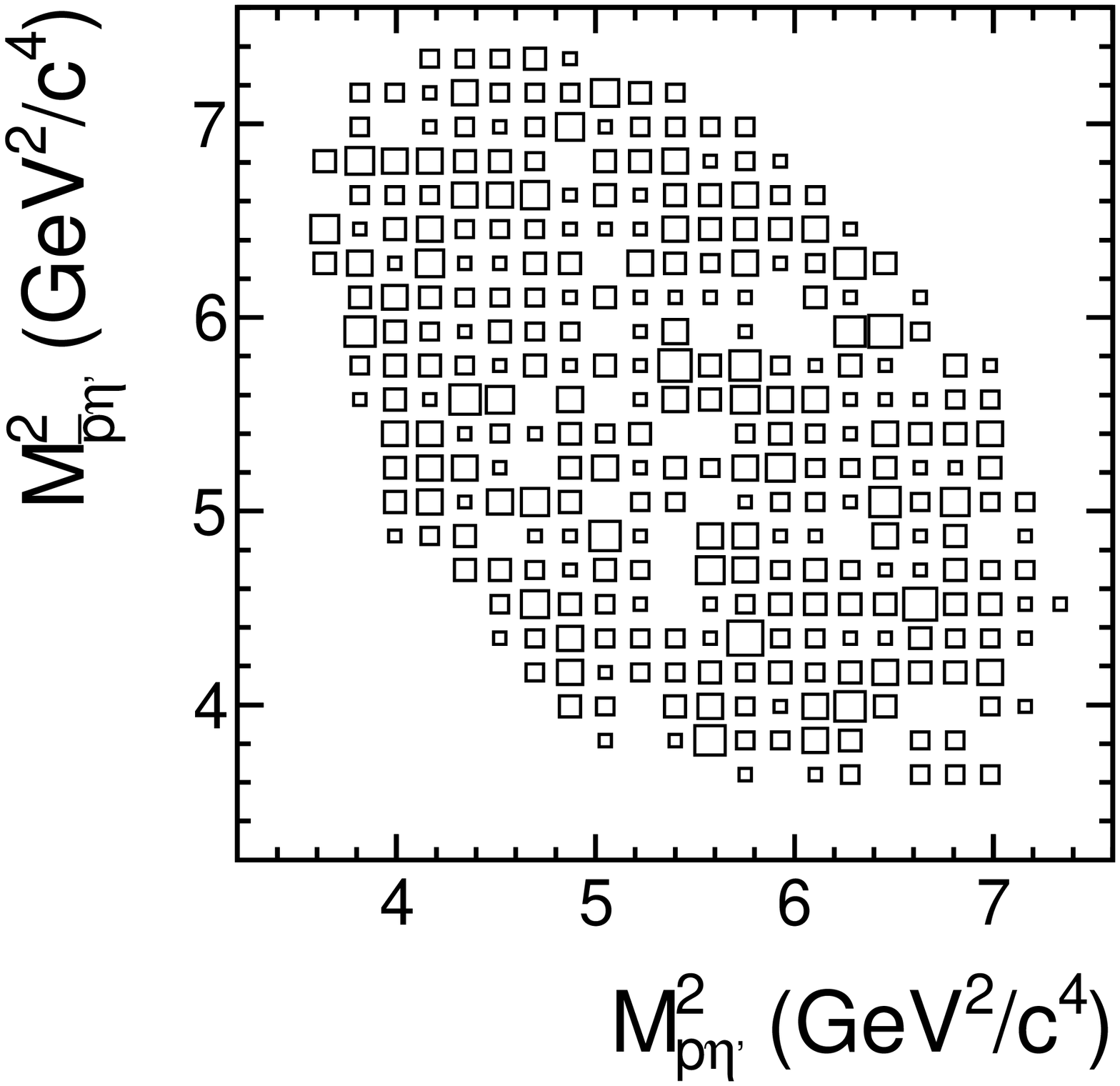}
    \put(80, 80){(a)}
  \end{overpic}
  \begin{overpic}[width=0.23\textwidth]{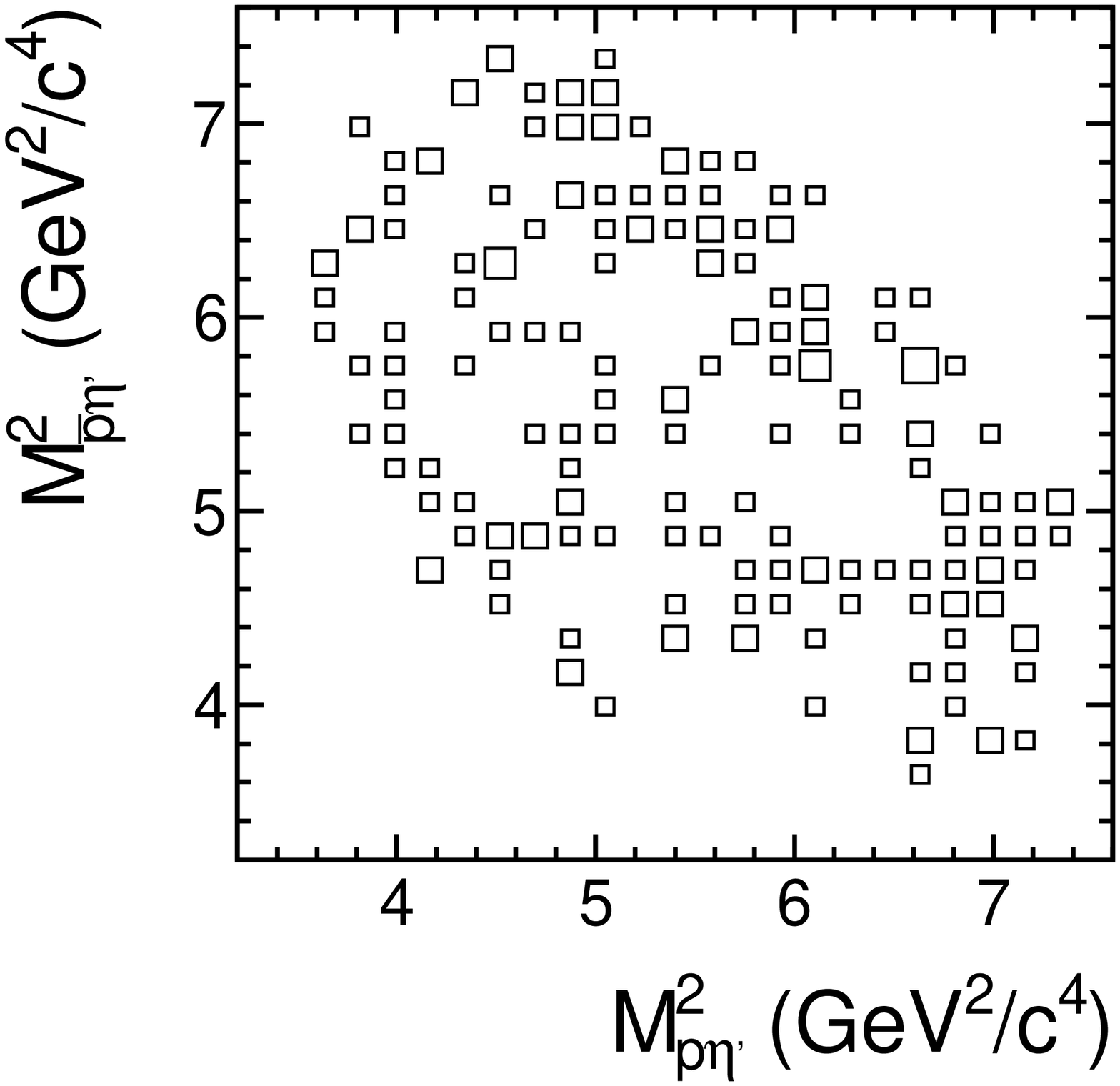}
    \put(80, 80){(b)}
  \end{overpic}
  \begin{overpic}[width=0.23\textwidth]{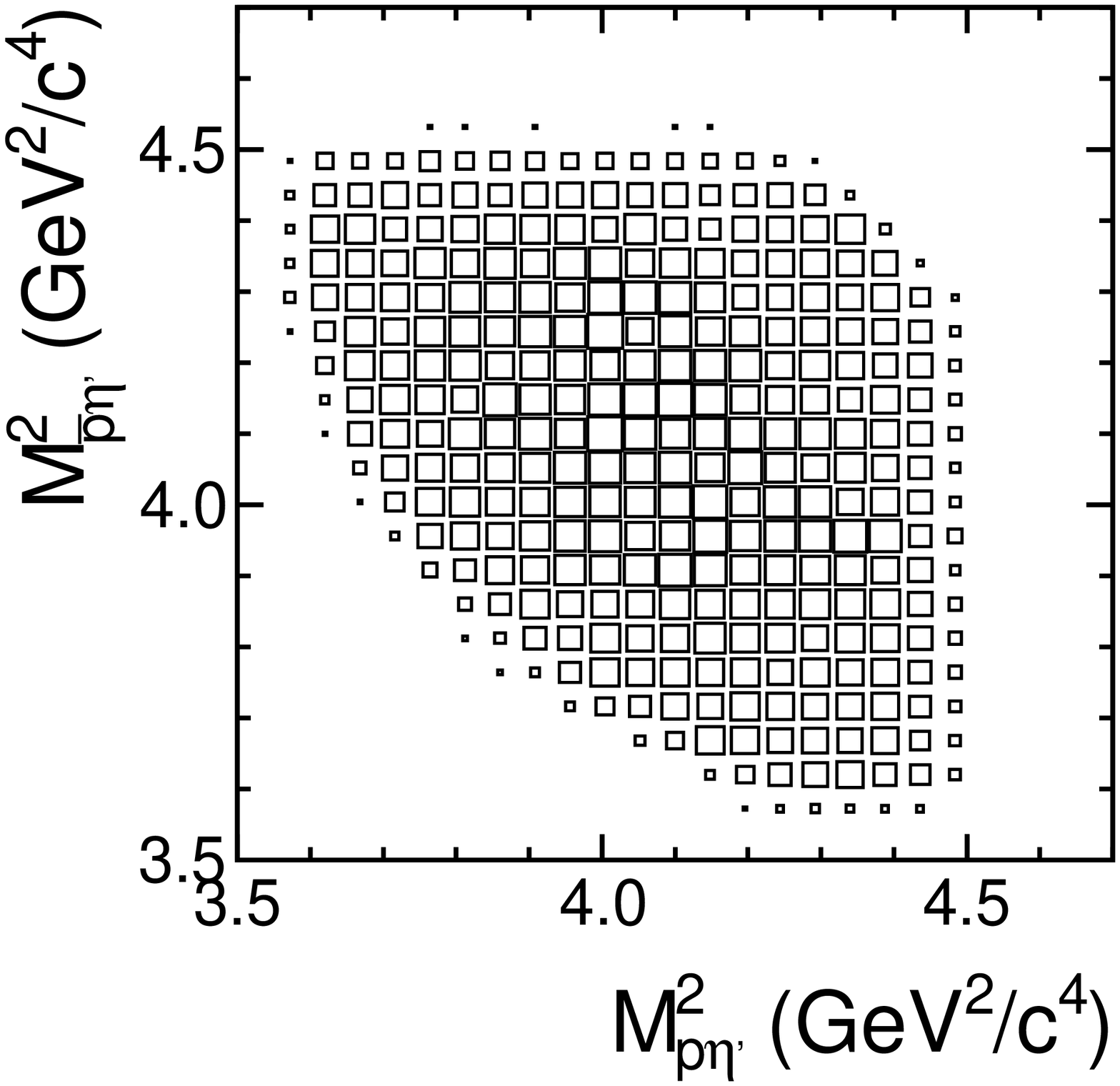}
    \put(80, 80){(c)}
  \end{overpic}
  \begin{overpic}[width=0.23\textwidth]{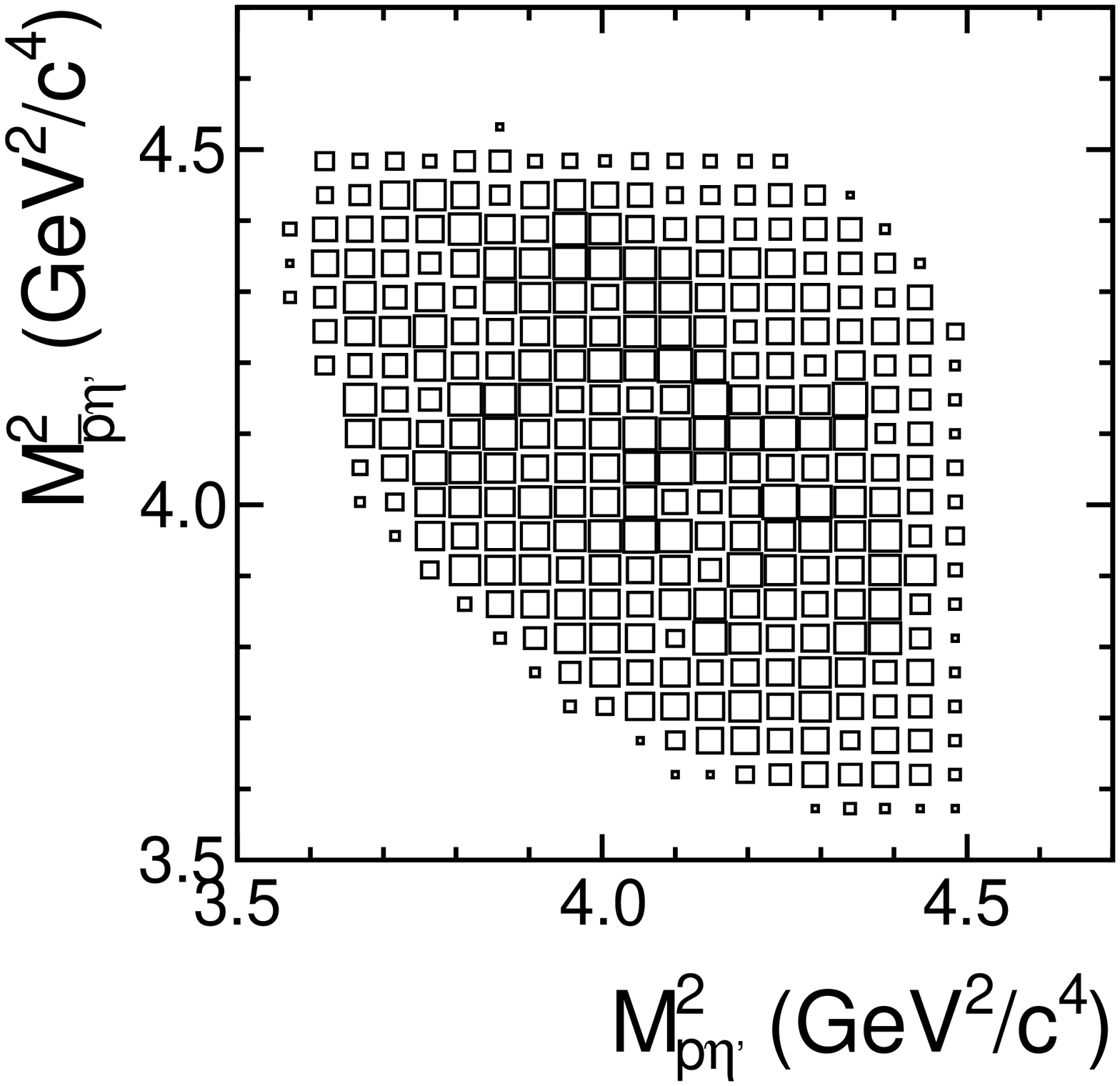}
    \put(80, 80){(d)}
  \end{overpic}
	\caption{Dalitz plots of $\psip\to\p\pb\etap$ with $\etap\to\gamma\pip\pim$ (a), $\psip\to\p\pb\etap$ with $\etap\to\eta\pip\pim$ (b), $\jpsi\to\p\pb\etap$ with $\etap\to\gamma\pip\pim$ (c), and $\jpsi\to\p\pb\etap$ with $\etap\to\eta\pip\pim$ (d).}
	\label{fig-dalitz}
\end{figure}

\par
\begin{figure*}[hbtp]
  \centering
  \begin{overpic}[width=0.23\textwidth]{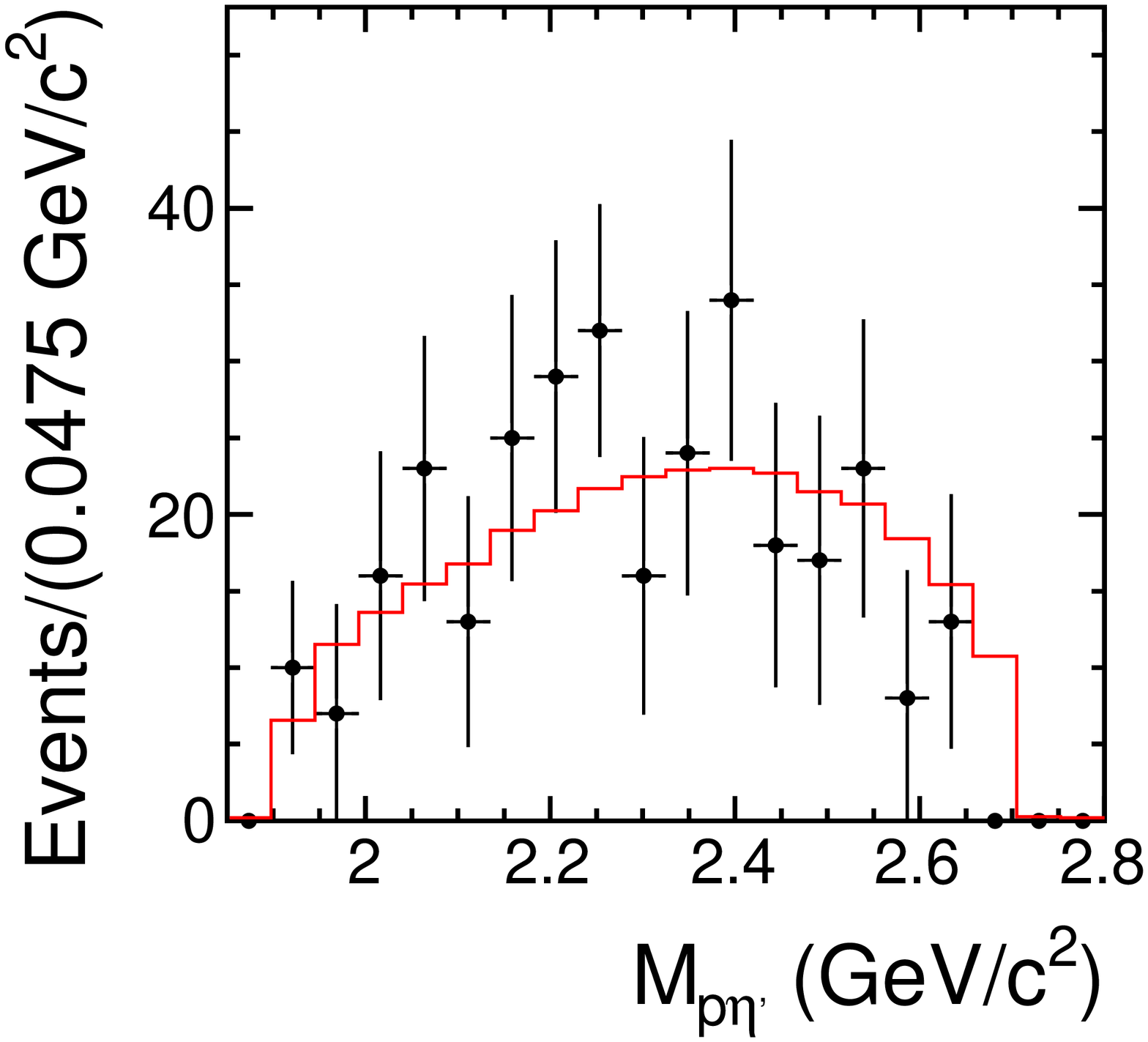}
    \put(80, 80){(a)}
  \end{overpic}
  \begin{overpic}[width=0.23\textwidth]{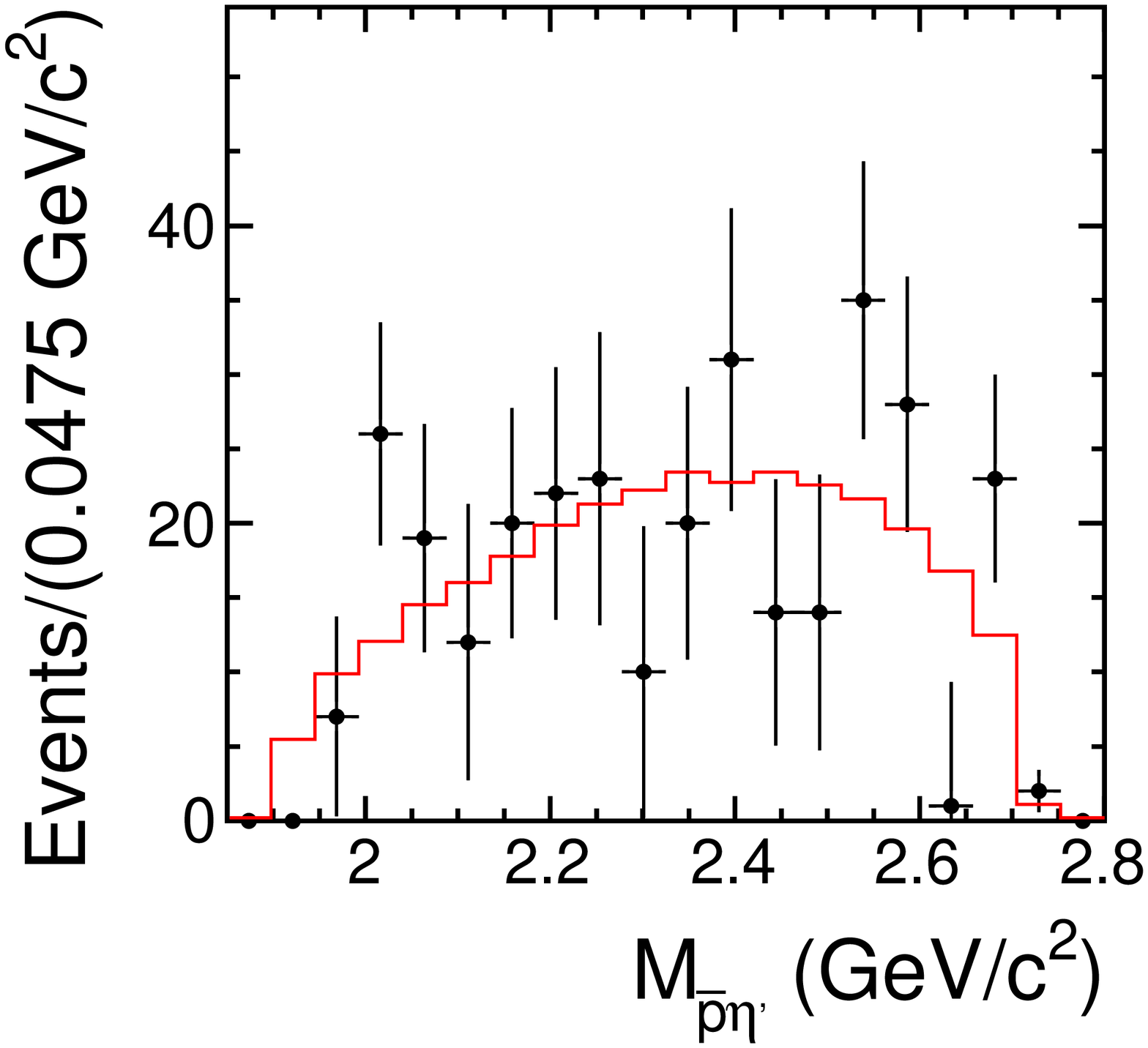}
    \put(80, 80){(b)}
  \end{overpic}
  \begin{overpic}[width=0.23\textwidth]{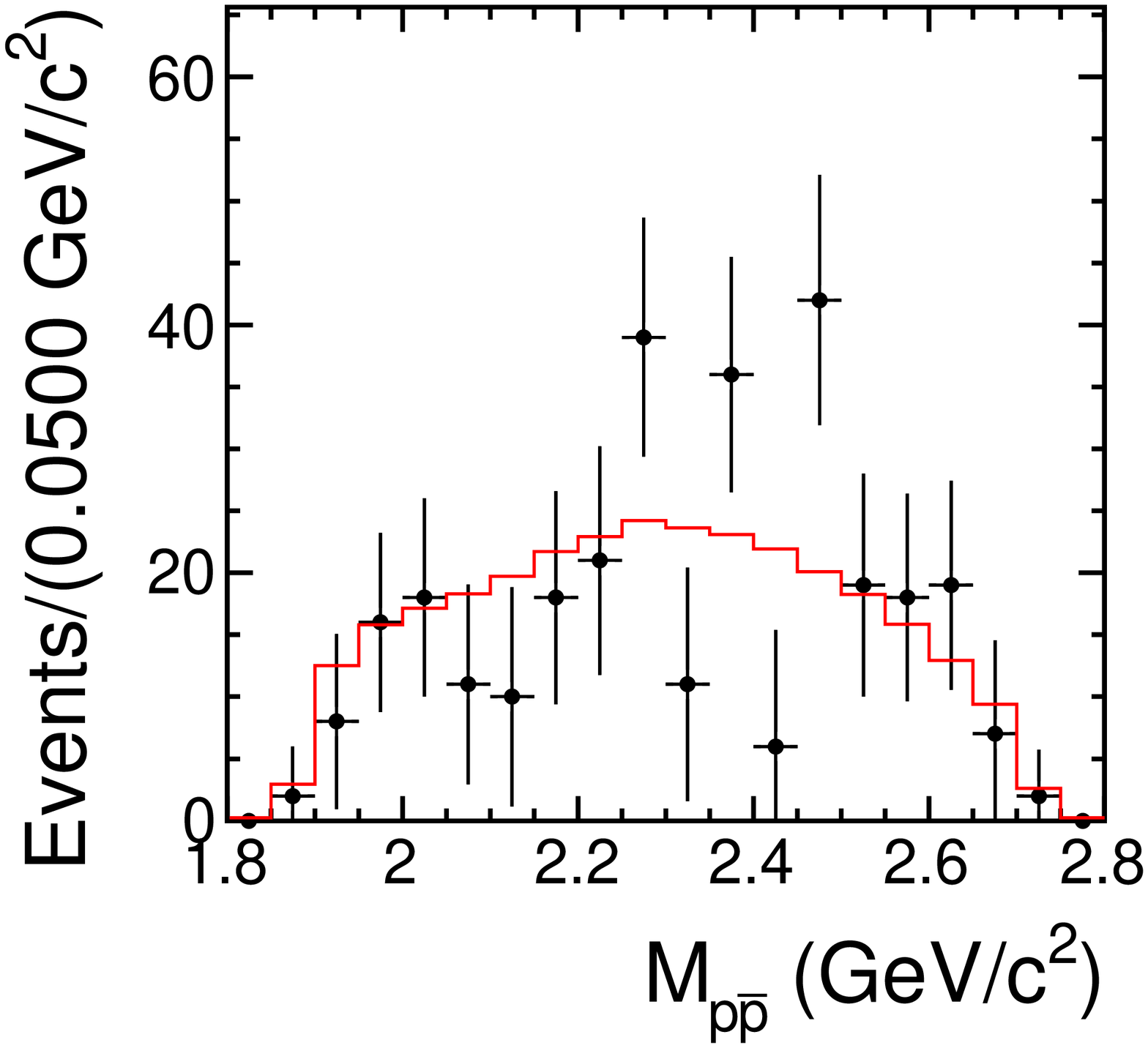}
    \put(80, 80){(c)}
  \end{overpic} \\
  \begin{overpic}[width=0.23\textwidth]{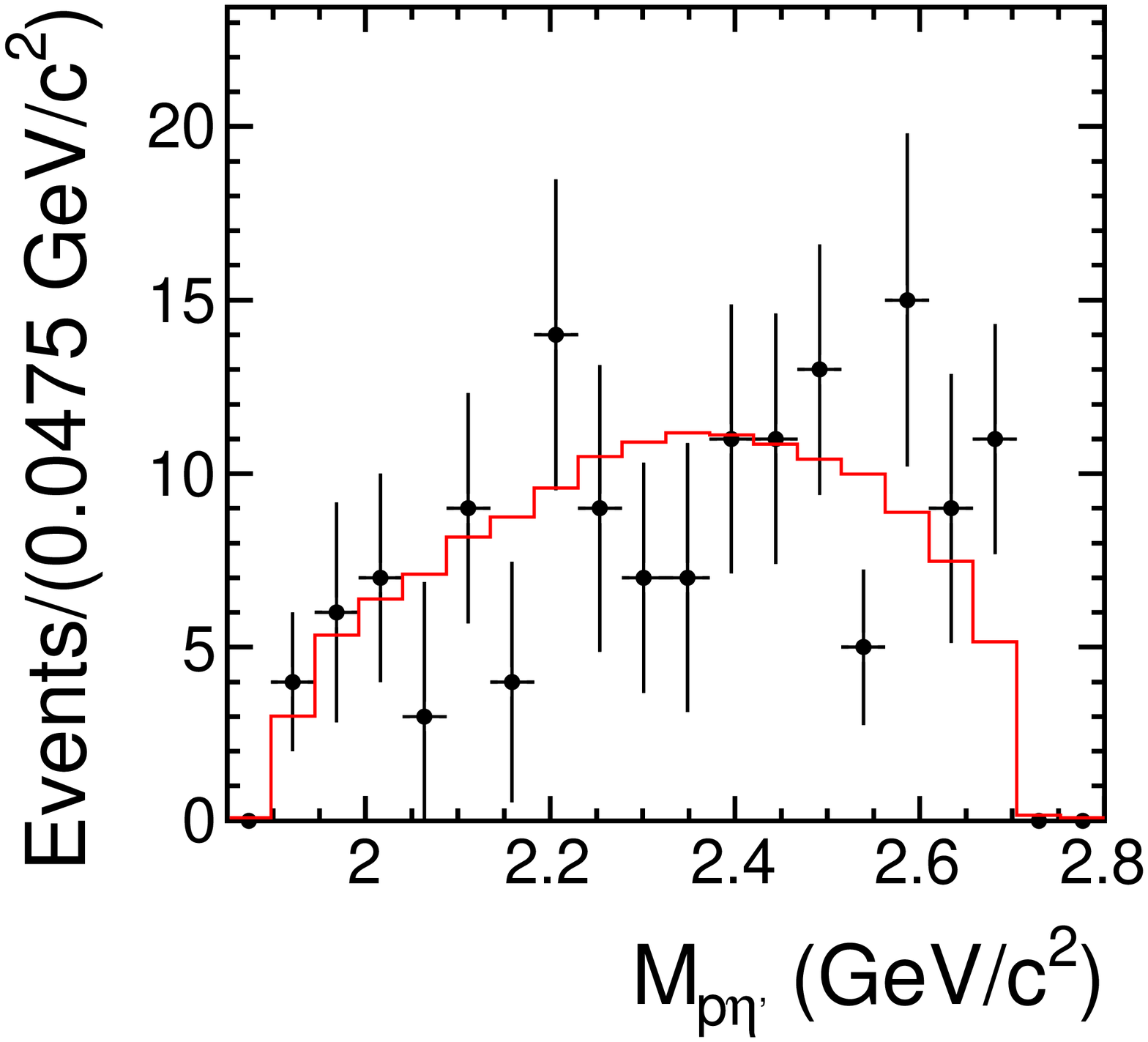}
    \put(80, 80){(d)}
  \end{overpic}
  \begin{overpic}[width=0.23\textwidth]{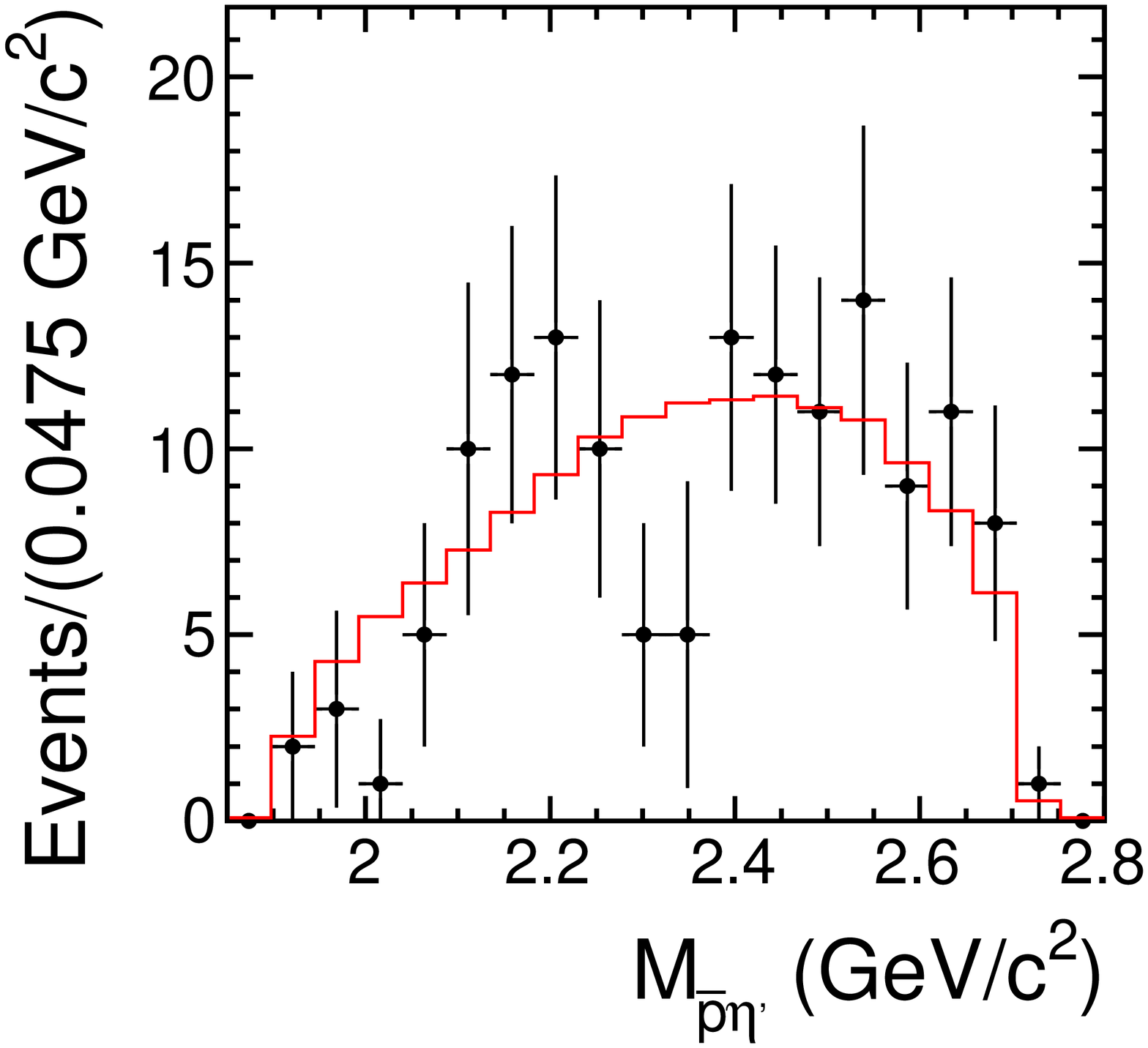}
    \put(80, 80){(e)}
  \end{overpic}
  \begin{overpic}[width=0.23\textwidth]{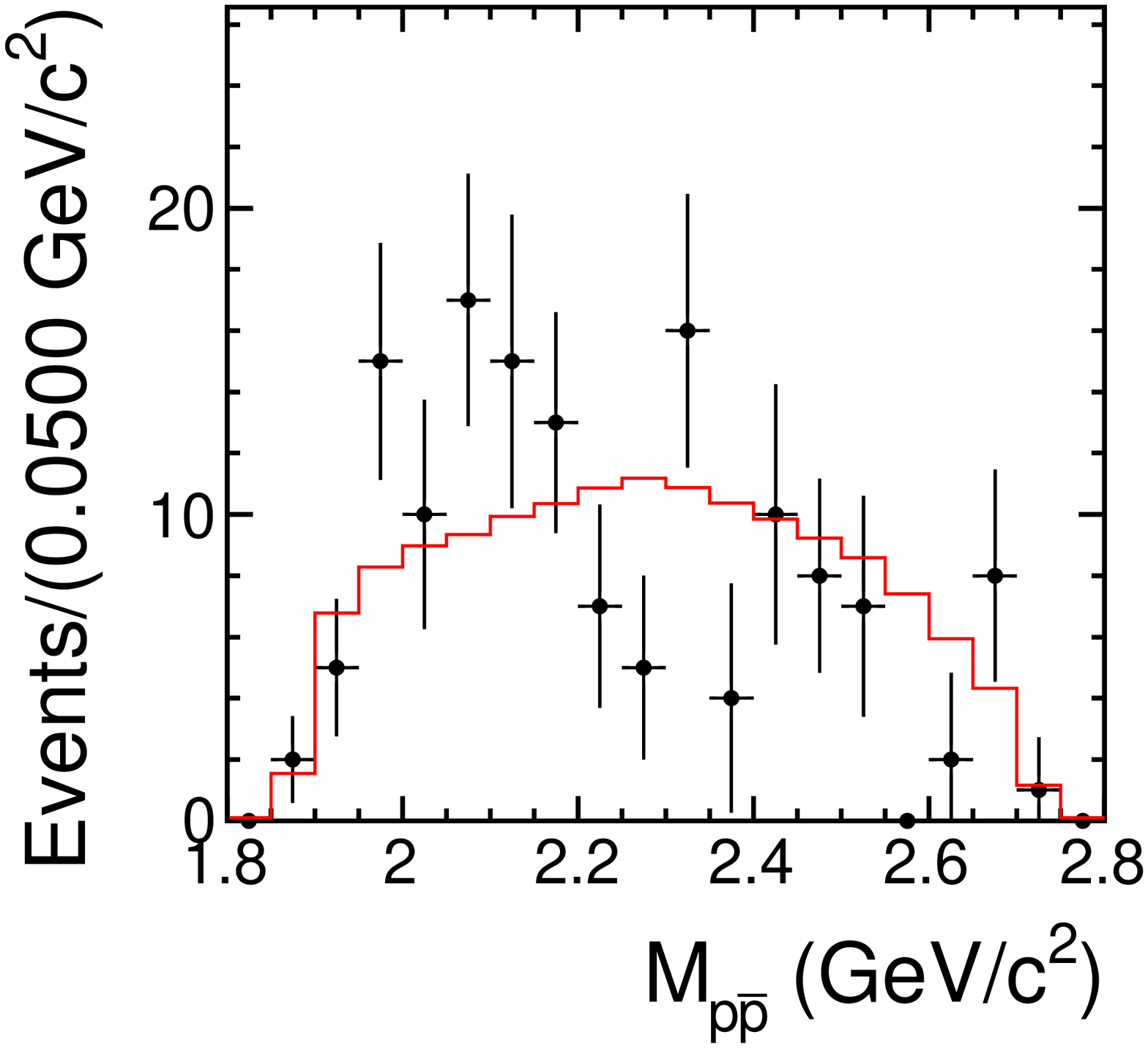}
    \put(80, 80){(f)}
  \end{overpic}
  \caption{Invariant mass distributions of $p\etap$ (a), $\pb\etap$ (b), and $p\pb$ (c) for $\psip \to p \pb \etap$ with $\etap \to \gamma \pip \pim$, and those of $p\etap$ (d), $\pb\etap$ (e), and $p\pb$ (f) for $\psip \to p \pb \etap$ with $\etap \to \pip \pim \eta$. The dots with error bars are data with background subtracted, and the red lines are the corresponding signal MC.}
	\label{fig-dalitz-psip-1d}
\end{figure*}

\begin{figure*}[hbtp]
  \centering
  \begin{overpic}[width=0.23\textwidth]{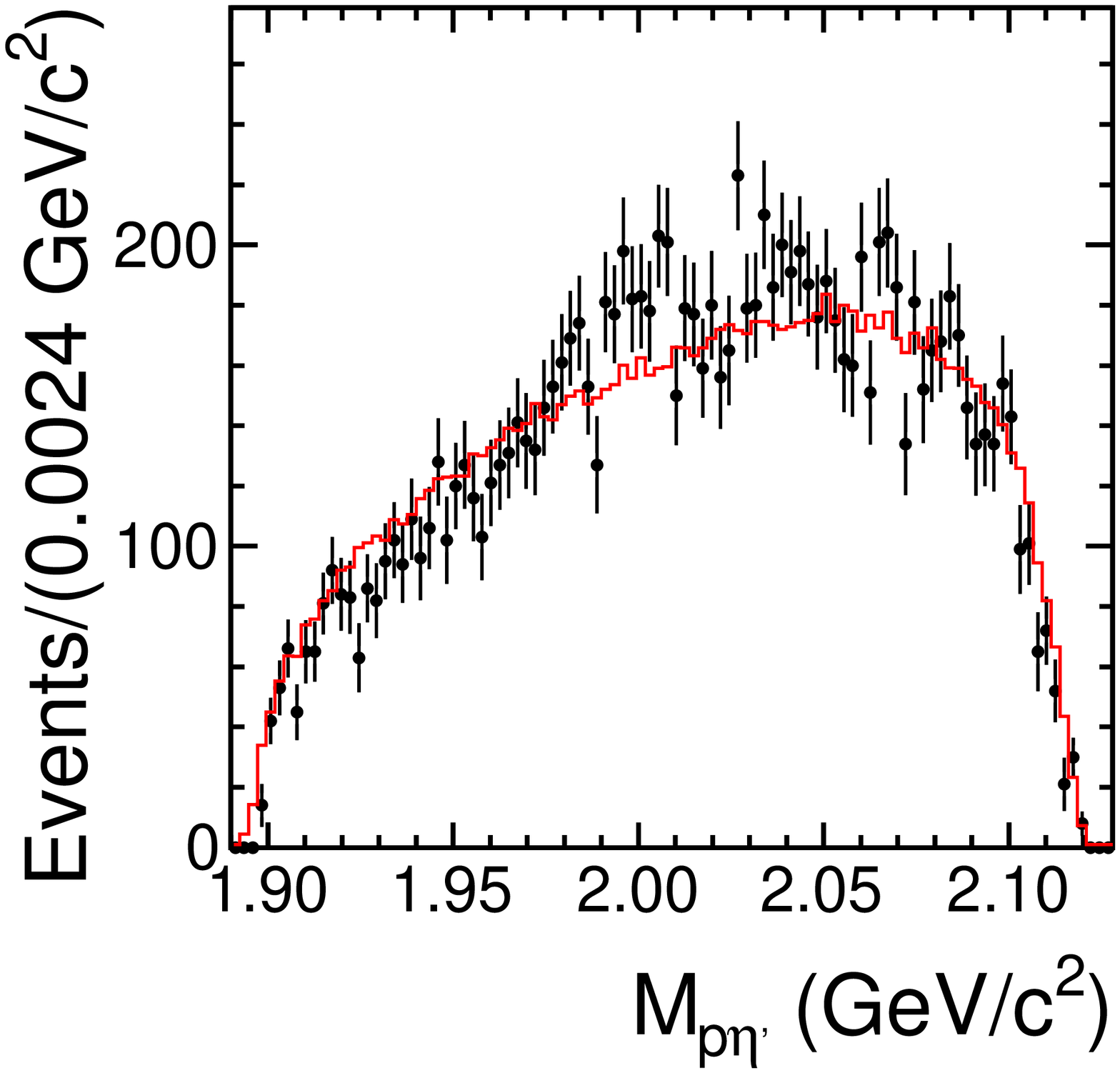}
    \put(80, 80){(a)}
  \end{overpic}
  \begin{overpic}[width=0.23\textwidth]{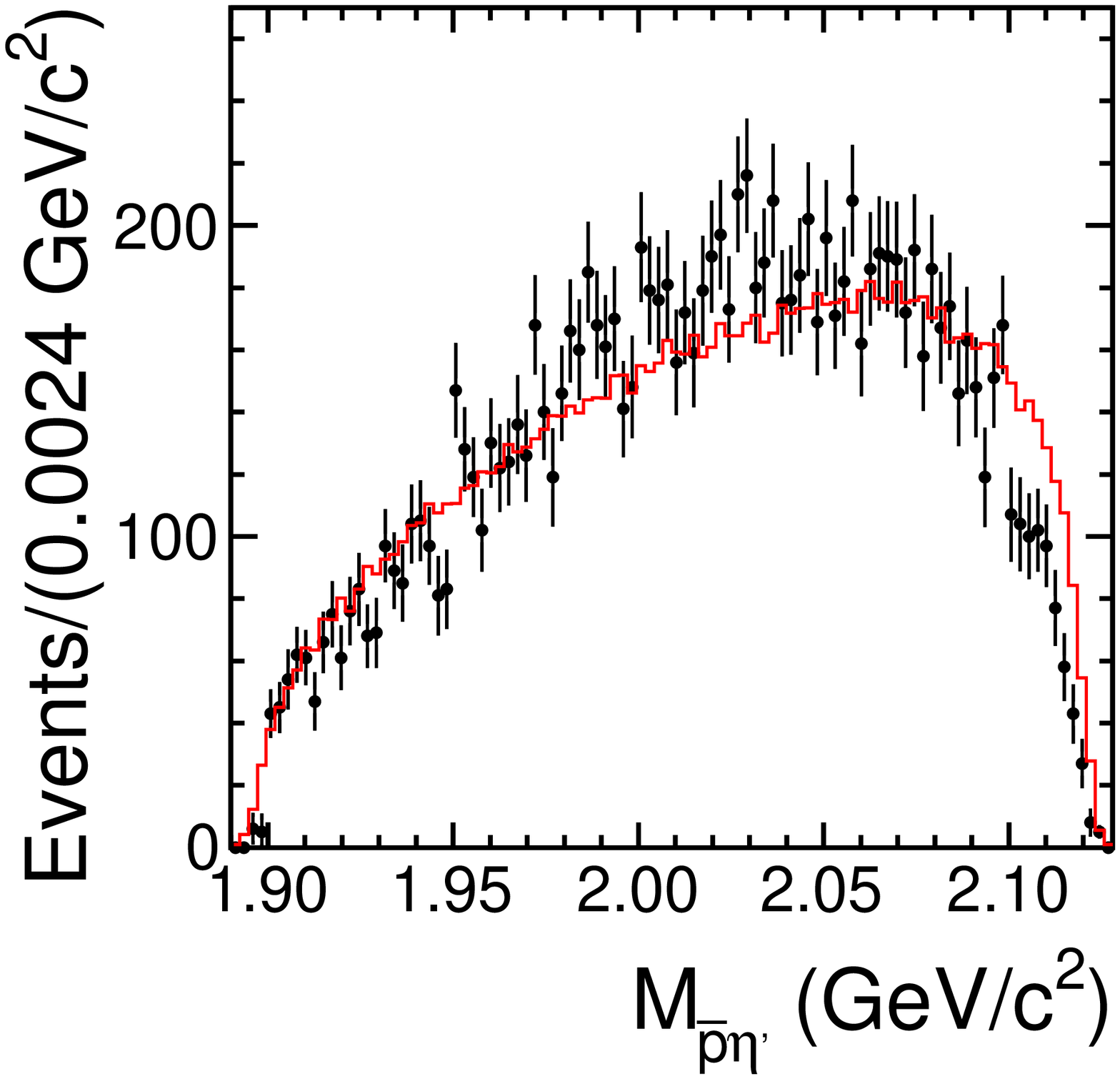}
    \put(80, 80){(b)}
  \end{overpic}
  \begin{overpic}[width=0.23\textwidth]{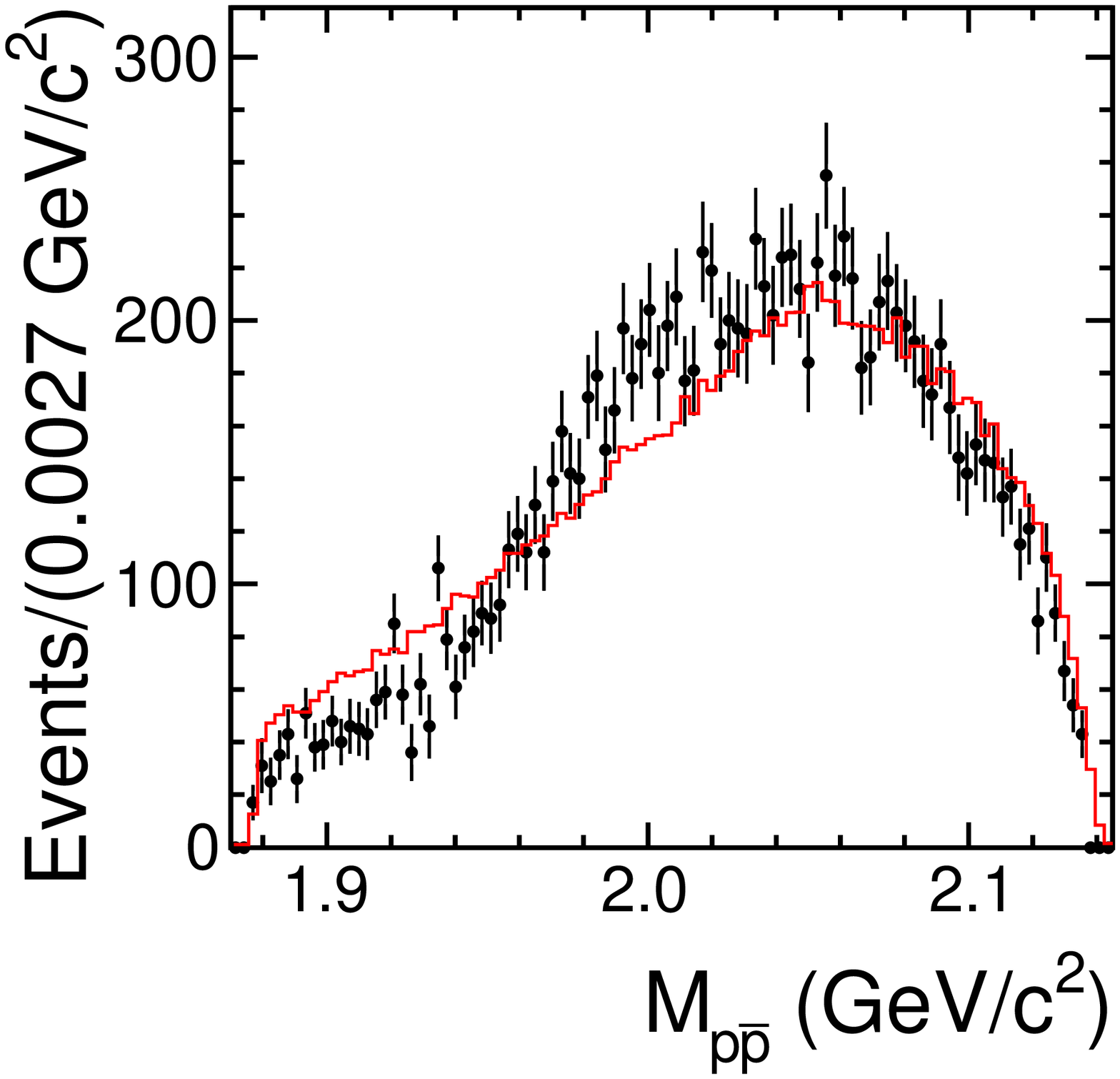}
    \put(80, 80){(c)}
  \end{overpic} \\
  \begin{overpic}[width=0.23\textwidth]{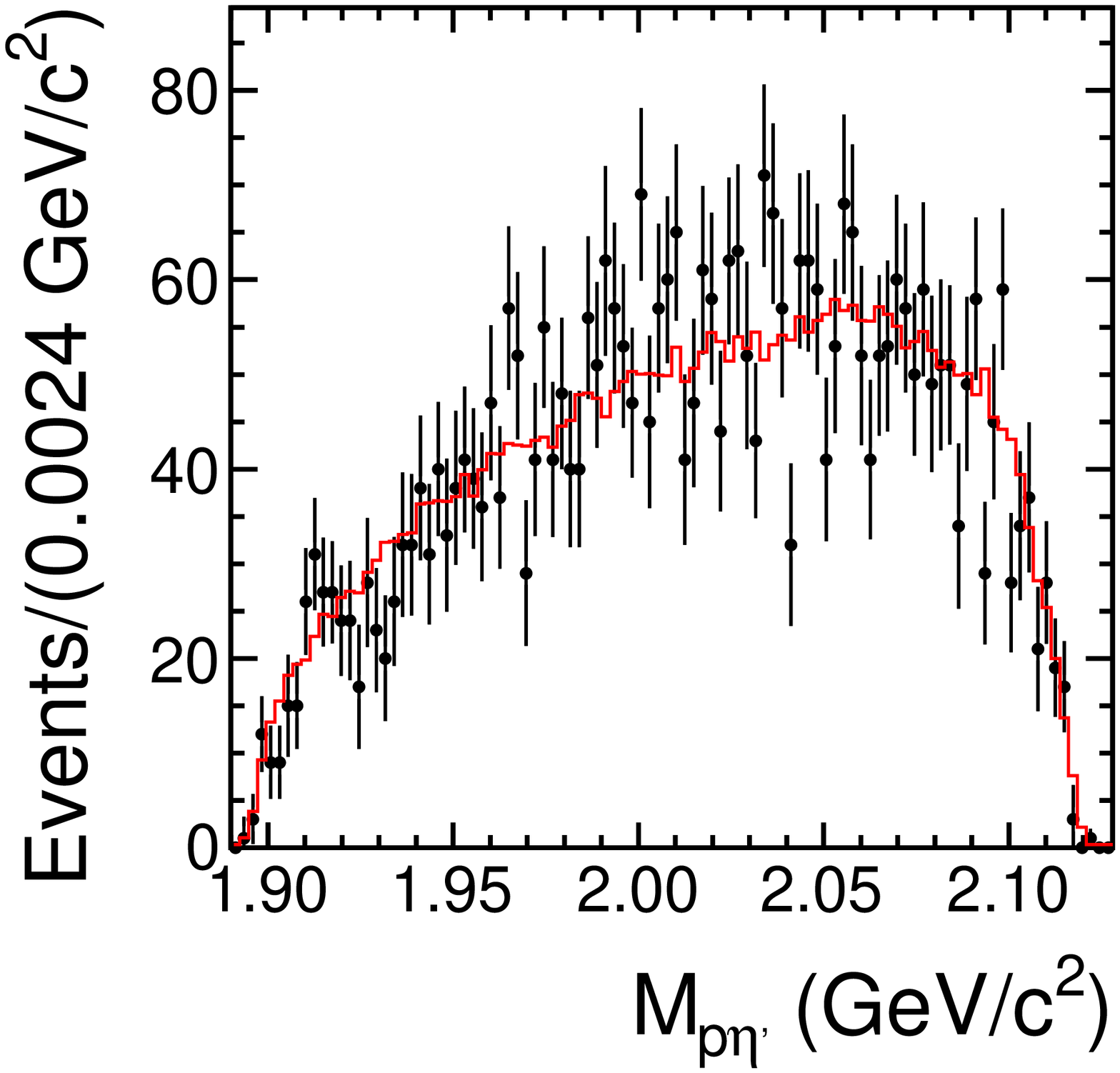}
    \put(80, 80){(d)}
  \end{overpic}
  \begin{overpic}[width=0.23\textwidth]{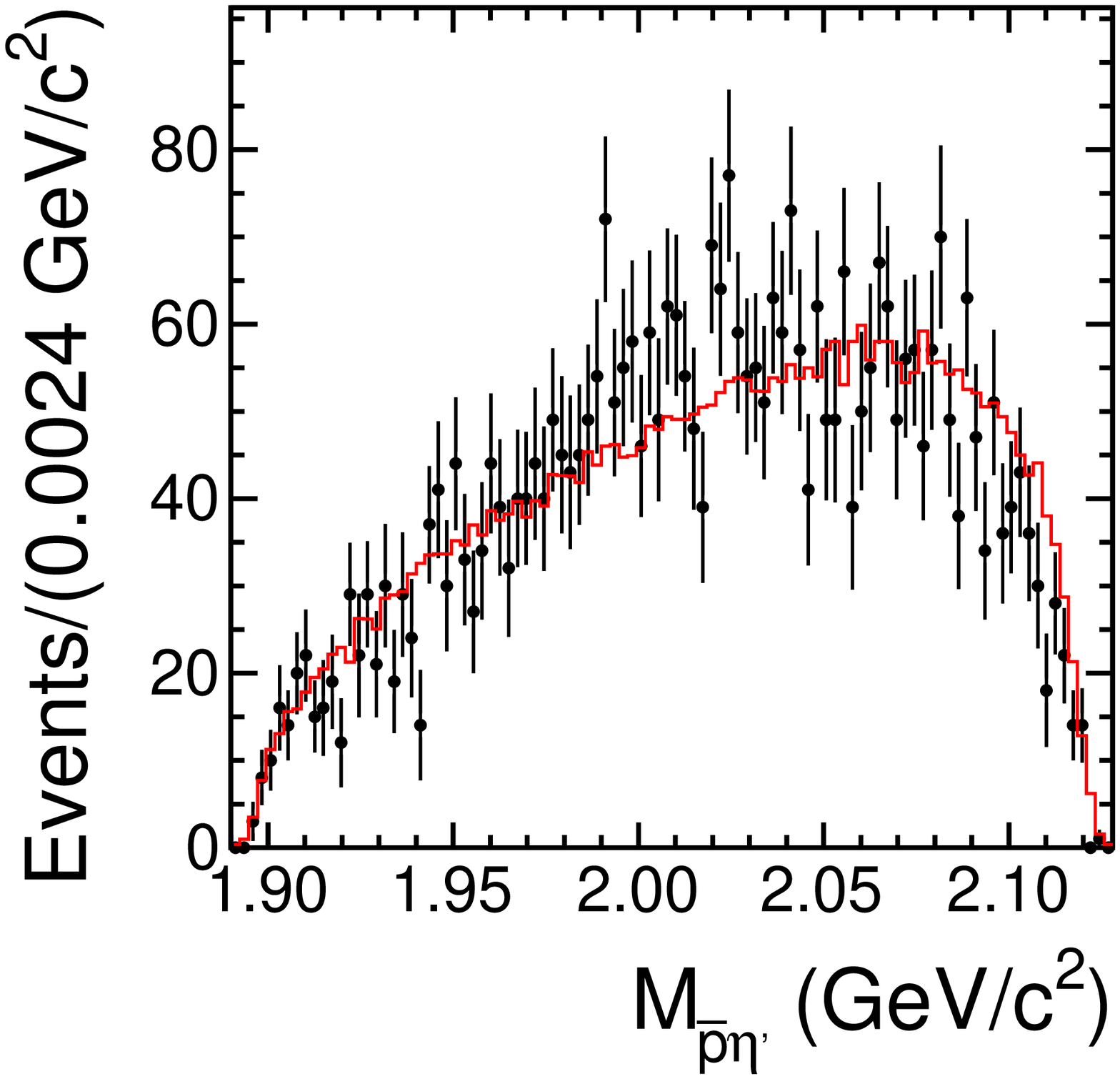}
    \put(80, 80){(e)}
  \end{overpic}
  \begin{overpic}[width=0.23\textwidth]{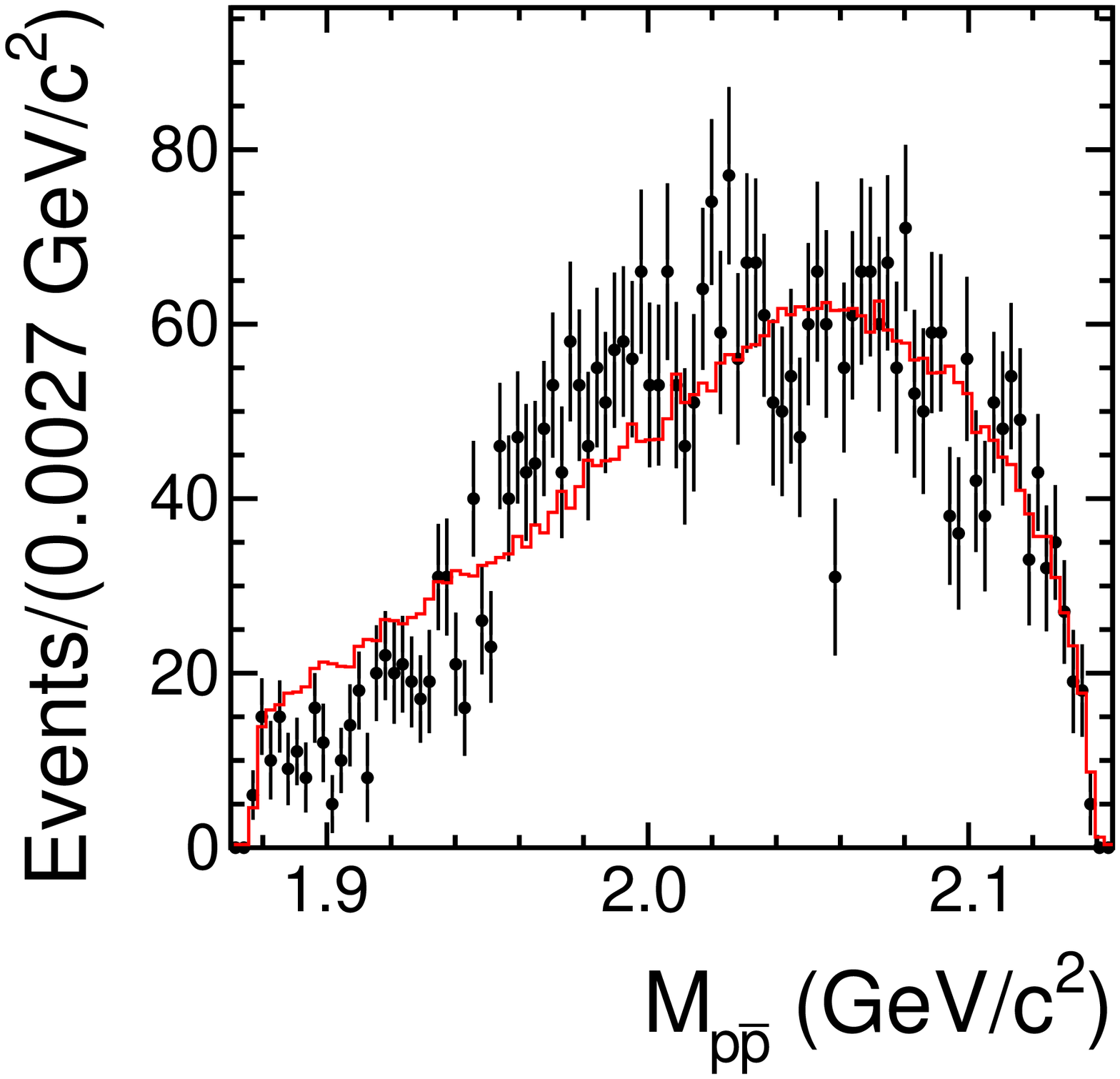}
    \put(80, 80){(f)}
  \end{overpic}
	\caption{Invariant mass distributions of $p\etap$ (a),
          $\pb\etap$ (b), and $p\pb$ (c) for $\jpsi \to p \pb \etap$
          with $\etap \to \gamma \pip \pim$, and those of $p\etap$
          (d), $\pb\etap$ (e), and $p\pb$ (f) for $\jpsi \to p \pb
          \etap$ with $\etap \to \pip \pim \eta$. The dots with error
          bars show background subtracted data, and the red lines are
          the corresponding distributions from signal MC.}
	\label{fig-dalitz-jpsi-1d}
\end{figure*}

 The $M_{\gamma\pip\pim}$ and $M_{\eta\pip\pim}$ distributions of the events that pass all selection criteria are shown in Fig.~\ref{fig-fit-invm-etap}. Peaks originating from $\etap$ decays are observed. Figure~\ref{fig-dalitz} shows the Dalitz plots of the events in the $\etap$ signal region, and Figs.~\ref{fig-dalitz-psip-1d} and \ref{fig-dalitz-jpsi-1d} show the invariant mass projections, where the side band backgrounds have been subtracted. Based on these plots, no obvious intermediate structures in invariant mass of $p\etap$, $\pb\etap$, or $p\pb$ are observed.

\section{SIGNAL YIELDS AND BRANCHING FRACTIONS}

\par To determine the branching fractions, simultaneous unbinned
maximum likelihood fits to the $\gamma\pip\pim$ and $\eta\pip\pim$ invariant mass spectra are performed for the $\psip$ data and for the $\jpsi$ data. The signal shape is represented by the MC-simulated $\etap$ mass distribution, convolved with a Gaussian function with free mean and width to account for the mass and resolution difference between data and MC simulation. The background is parameterized as a second-order Chebyshev polynomial with free parameters. In the simultaneous fit, the ratio of the number of $\etap \to \gamma \pip \pim$ events to that of $\etap \to \eta \pip \pim$ events is fixed to  $\frac{\br(\etap \to \gamma \pip \pim) \cdot \epsilon_{\etap \to \gamma \pip \pim}}{\br(\etap \to \pip \pim \eta) \cdot \br(\eta \to \gamma \gamma) \cdot \epsilon_{\etap \to \pip \pim \eta}}$, where $\epsilon_{\etap \to \gamma \pip \pim}$ and $\epsilon_{\etap \to \pip \pim \eta}$ are the global efficiencies for each $\etap$ decay mode. Due to differences in tracking and PID efficiencies between data and MC simulation for protons and anti-protons, the MC-determined global efficiencies are corrected by multiplying factors 1.030 (1.038) and 0.980 (0.984) for tracking and PID, respectively, for $\psip\to\p\pb\etap$ ($\jpsi\to\p\pb\etap$). These correction factors are ratios of efficiencies between data and MC simulation obtained by studying the control samples $\bpsi \to\p\pb\pip\pim$, where the efficiencies are weighted according to the distributions of transverse momentum (for tracking) or momentum (for PID) of protons and anti-protons.

\par The results of the fits are listed in Table~\ref{tab-reslist} and shown in Fig.~\ref{fig-fit-invm-etap}. The goodness of the fit is $\chisq / {\rm ndf} = 51.50 / 44 = 1.17$ for $\psip \to \p \pb \etap$ with $\etap \to \gamma \pip \pim$, $37.85 / 44=0.86$ for $\psip \to \p \pb \etap$ with $\etap \to \eta \pip \pim$, $234.75 / 194 = 1.21$ for $\jpsi \to \p \pb \etap$ with $\etap \to \gamma \pip \pim$, and $205.66 / 194 = 1.06$ for $\jpsi \to \p \pb \etap$ with $\etap \to \eta \pip \pim$. The resolution difference between data and MC simulation is 1 MeV in each decay mode. The branching fractions are determined to be
\begin{align*}
  \br(\psip\to\p\pb\etap)&=(1.10\pm0.10\pm0.08)\times10^{-5}, \\
  \br(\jpsi\to\p\pb\etap)&=(1.26\pm0.02\pm0.07)\times10^{-4}.
\end{align*}
Here the first uncertainties are statistical and the second ones systematic, as discussed in Sec.~\ref{sec:sys}. As a cross check, we also fit these channels separately, and Table~\ref{tab-reslist} lists the signal yields, the selection efficiencies, and the branching fractions obtained for each decay mode. The branching fractions obtained from the simultaneous and separate fits are consistent with each other.

\begin{table*}[hbtp]
  \centering
  \caption{Results from separate and simultaneous fits for each decay, $N_{\rm sig}$ is the number of signal events, $\epsilon$ is the selection efficiency, and $\br$ is the branching fraction of $\psip\to p\pb \etap$ or $\jpsi\to p\pb \etap$.}
  \begin{tabular}{l|ccc|ccc}
    \hline
    \hline
      &\multicolumn{3}{c|}{$\psip \to\p\pb\etap$} &\multicolumn{3}{c}{$\jpsi\to \p \pb\etap$} \\
    \hline
    Category &$N_{\rm sig}$ &$\epsilon$~(\%) &$\br$~($10^{-5}$) &$N_{\rm sig}$ &$\epsilon$~(\%) &$\br$~($10^{-4}$) \\
    \hline
    $\etap\to\gamma\pip\pim$ &$337\pm29$ &22.6 &$1.12\pm0.10\pm0.10$ &$12390\pm138$ &25.5 &$1.27\pm0.02\pm0.08$ \\
    $\etap\to\eta\pip\pim$   &$154\pm14$ &18.7 &$1.07\pm0.10\pm0.08$ &$3931\pm74$   &14.3 &$1.24\pm0.03\pm0.10$ \\
    Simultaneous fit         & --        & --   &$1.10\pm0.10\pm0.08$ & --           & --   &$1.26\pm0.02\pm0.07$ \\
    \hline
    \hline
  \end{tabular}
  \label{tab-reslist}
\end{table*}

\section{SYSTEMATIC UNCERTAINTIES}
\label{sec:sys}
\par The systematic uncertainties mainly come from the MDC tracking, photon and $\eta$ reconstruction, PID, kinematic fit, mass windows, branching fractions of the decay modes used to reconstruct the $\etap$, the number of {\boldmath $\psi$} decays, fitting procedure, and the physics model used to determine the efficiency. All the contributions are given in Table~\ref{tab-sys-err}. The overall systematic uncertainties are obtained by adding all systematic uncertainties, taking the correlations into account.

\par The uncertainty in the MDC tracking efficiency for each pion is estimated with the control sample $\psip\to\pip\pim\jpsi$, and a 1.0\% systematic
uncertainty per pion is obtained~\cite{tracking_pion}. This gives a total of 2.0\% for each decay mode.
The tracking efficiencies of protons and anti-protons are studied with the control sample $\bpsi\to\p\pb\pip\pim$.
The MC efficiencies for the signal processes are corrected using the results from the control samples,
and the uncertainties of the tracking differences between data and MC simulation are taken as the systematic uncertainties,
which are 0.7\% (1.0\%) per proton (anti-proton) for the $\psip$ decays, and 0.9\% (1.0\%) per proton (anti-proton) for the $\jpsi$ decays.
Assuming the uncertainties of proton and anti-proton are totally correlated, the tracking uncertainty of a proton--anti-proton pair is 1.7\% (1.9\%) for the $\psip$ ($\jpsi$) data sample.
The tracking efficiencies of pion and proton are independent, and the total tracking uncertainty is determined to be 2.7\% (2.8\%)  for the $\psip$ ($\jpsi$) samples.

\par The uncertainty in the photon reconstruction is studied by using the control sample $\jpsi\to\rho^{0}\pi^{0}$, and a 1.0\% systematic uncertainty
is estimated for each photon~\cite{photon_rec}. The uncertainty of the $\eta$ reconstruction from $\gamma\gamma$ final state is 1.0\% per $\eta$,
as determined from a high purity control sample of $\jpsi\to\p\pb\eta$~\cite{eta-rec}.

\par The uncertainty in the PID efficiency for pions is estimated to be 1.0\% per pion~\cite{pid_pion}, and
the total PID uncertainty for two pions is 2.0\% for each decay mode.
The efficiencies of the proton and the anti-proton identification are studied with the control samples $\bpsi\to\p\pb\pip\pim$.
The MC efficiencies for the signal processes are corrected using the results from the control samples,
and the uncertainties of the corrections are taken as the systematic
uncertainties, which are 0.5\% (0.6\%) per proton (anti-proton) for the $\psip$, and 0.6\% (0.8\%) per proton (anti-proton) for the $\jpsi$ samples.
Assuming the uncertainties of proton and anti-proton are totally correlated, the PID uncertainty of a proton and anti-proton pair is calculated to be 1.1\%(1.4\%) for the $\psi(3686)$ ($J/\psi$) data samples.
The PID efficiencies of pion and proton are independent, and the total PID uncertainty is determined to be 2.3\% (2.5\%) for the $\psip$ ($\jpsi$) samples.

\par The uncertainty associated with the kinematic fit is estimated by comparing the efficiencies with or without a
helix parameter correction applied to simulated data~\cite{bib23}. Control samples $\jpsi\to\p K^{-}\bar{\Lambda}+c.c.$ and $\psip\rightarrow K^{+}K^{-}\pi^{+}\pi^{-}$~\cite{bib24} are used to obtain the correction to the track helix parameters. The uncertainties due to the kinematic fit are determined to be 1.9\%, 2.4\%, 1.7\%, and 2.3\%, for $\psip\to\p\pb\etap$ with $\etap\to\gamma\pip\pim$,
and $\psip\to\p\pb\etap$ with $\etap\to\eta\pip\pim$, $\jpsi\to\p\pb\etap$ with $\etap\to\gamma\pip\pim$, and $\jpsi\to\p\pb\etap$ with $\etap\to\eta\pip\pim$, respectively.

\par The uncertainty due to the mass windows used to veto background events originates from the differences in the mass resolutions between data and MC simulation.
We repeat the analysis by enlarging or reducing the mass window. The largest difference is used as an estimate of the corresponding systematic uncertainty. The $\etap$ signal region and sideband regions are not used to veto the background events, so they have no effect on the branching fraction determination. The uncertainties due to different mass windows are considered to be independent, so we add them in quadrature. For the decay $\psip\to\p\pb\etap$ with $\etap\to\gamma\pip\pim$ ($\etap\to\eta\pip\pim$), the uncertainty is 4.9\% (1.8\%). The uncertainty for the mass-window selection for $\jpsi\to\p\pb\etap$ is found negligible.

\par The systematic uncertainties due to the branching fractions of the subsequent $\etap$ and $\eta$ decays are 1.7\% for decays $\etap\to\gamma\pip\pim$ and $\etap\to\eta\pip\pim\to\gamma\gamma\pip\pim$~\cite{pdg_2014}. The numbers of $\psip$ and $\jpsi$ events have been estimated via inclusive hadronic events with relative uncertainties of 0.7\%~\cite{psip_num} and 0.6\%~\cite{jpsi_num}, respectively.

\par The fit range, signal shape, and background shape are considered as the sources of the systematic uncertainty related with the fit procedure.
In the nominal fit, the mass range is $[0.90,1.04]$~GeV/$c^{2}$, and we repeat the fit by changing the range by $\pm$10~MeV/$c^{2}$. The largest change in the final result is taken as the uncertainty due to the fit range, which is 1.4\% and 0.6\% for $\psip\to\p\pb\etap$ and $\jpsi\to\p\pb\etap$, respectively. For the signal shape, we change the nominal shape to a double-Gaussian or a Breit-Wigner function convolved with a Gaussian function, and the largest difference from the nominal result is taken as the uncertainty of the signal shape, which is 3.9\% and 1.2\% for $\psip\to\p\pb\etap$ and $\jpsi\to\p\pb\etap$, respectively. We replace the background shape with a first-order Chebyshev or second-order polynomial function, the largest differences from the nominal result, 2.7\% and 1.3\% for $\psip\to\p\pb\etap$ and $\jpsi\to\p\pb\etap$, respectively, are taken as systematic uncertainties.

\par The signal MC sample is generated assuming pure phase space distribution, in which possible intermediate states and non-flat angular distributions are ignored.
Although no strong structure is visible in the Dalitz plots shown in Fig.~\ref{fig-dalitz}, the phase space MC does not provide a very good description of the data, as shown in Fig.~\ref{fig-dalitz-psip-1d} and Fig.~\ref{fig-dalitz-jpsi-1d}, which results in a large systematic uncertainty, especially for the $\jpsi$ decay modes.
For $\psip\to\p\pb\etap$, since the statistics are limited, we weight each MC-generated phase space event by the $M_{p\bar{p}}$ distribution, and a difference of 1.5\% in the efficiency between the nominal and weighted MC samples is taken as the uncertainty. For $\jpsi\to\p\pb\etap$, we re-generate signal MC events based on {\sc body3}~\cite{Ablikim:2014dqd}, a data-driven MC generator, and a difference of 3.4\% in the efficiencies is taken as the uncertainty.

\begin{table}[!h]
  \centering
  \small
  \caption{Summary of relative systematic uncertainties (in \%) in the branching fractions. The
    total systematic uncertainty is obtained by summing all the contributions from each source taking the correlations into account. Here I (II) is $\psip$ decay channel with $\etap\to \gamma \pip \pim$ ($\etap\to\eta\pip\pim$), and III (IV) is $\jpsi$ decay channel with $\etap\to \gamma \pip \pim$ ($\etap\to\eta\pip\pim$).}
  \begin{tabular}{lp{1cm}p{1cm}p{1cm}p{1cm}}
    \hline
    \hline
    Source &I  &II  &III  &IV  \\
    \hline
    Tracking &2.7 &2.7 &2.8 &2.8 \\
    Photon &1.0 &2.0 &1.0 &2.0 \\
    $\eta$ reconstruction &-- &1.0 &-- &1.0 \\
    PID &2.3 &2.3 &2.5 &2.5 \\
    Kinematic fit &1.9 &2.4 &1.7 &2.3 \\
    Mass window &4.9 &1.8 &-- &-- \\
    Branching fraction &1.7 &1.7 &1.7 &1.7 \\
    Number of {\boldmath $\psi$} events &\multicolumn{2}{c}{0.7} &\multicolumn{2}{c}{0.6} \\
    Fit range &\multicolumn{2}{c}{1.4} &\multicolumn{2}{c}{0.6} \\
    Signal shape &\multicolumn{2}{c}{3.9} &\multicolumn{2}{c}{1.2} \\
    Background shape &\multicolumn{2}{c}{2.7} &\multicolumn{2}{c}{1.3} \\
    Physics model &\multicolumn{2}{c}{1.5} &\multicolumn{2}{c}{3.4} \\
    Sum &\multicolumn{2}{c}{7.2} &\multicolumn{2}{c}{5.4} \\
    \hline
    \hline
  \end{tabular}
  \label{tab-sys-err}
\end{table}

\section{SUMMARY AND DISCUSSION}

\par Based on $4.48 \times 10^8$ $\psip$ decays, we observe for the first time $\psip\to\p\pb\etap$, and measure its branching fraction to be $(1.10\pm0.10\pm0.08)\times10^{-5}$. Based on $1.31 \times 10^{9}$ $\jpsi$ decays, we obtain the most accurate measurement so far of $\br(\jpsi\to\p\pb\etap)=(1.26\pm0.02\pm0.07)\times10^{-4}$.

\par With the measurement of $J/\psi \to p \bar{p}$~\cite{pdg_2014}, we determine the ratio of decay widths $\frac{\Gamma(J/\psi \to p \bar{p} \eta')}{\Gamma(J/\psi \to p \bar{p} )} = (5.94 \pm 0.35)\%$. It is larger than the nucleon pole contribution by two orders of magnitude according to Ref.~\cite{nucleon_pole}.  This implies the validity of the $N^*$ pole hypothesis. However, from the invariant mass distributions of $p \eta'$ and $\bar{p} \eta'$ in Fig.~\ref{fig-dalitz-jpsi-1d}, no distinctive structure is observed, which indicates that very broad intermediate $N^*$ states or other decay mechanisms are needed to explain the large ratio. Similarly, using the results in Ref.~\cite{psip_to_NN}, we determine the ratio $\frac{\Gamma(\psi(3686) \to p \bar{p} \eta')}{\Gamma(\psi(3686) \to p \bar{p} )} = (3.61 \pm 0.41)\%$, where common uncertainties have been cancelled. This ratio will helpful for future studies of the nucleon and $N^*$ pole contributions in $\psi(3686)$ baryonic decays.

\par Combining our result with the branching fractions of $\bpsi\to\p\pb\eta$ reported in Ref.~\cite{pdg_2014} and following the procedure described in Ref.~\cite{soft_pion_theorem}, we determine an $\eta-\etap$ mixing angle of $-24^{\circ}\pm11^{\circ}$ for the $\psip$ decays and
$-24^{\circ}\pm9^{\circ}$ for the $\jpsi$ decays. We observe that the two values are very similar, even though the uncertainties are in both cases very large. This might indicate a universal behavior of the $\eta-\etap$ mixing angle as expected.  These results are consistent with the QCD-inspired calculations $\theta_{\eta-\etap}=-(17^{\circ} \sim 10^{\circ})$~\cite{soft_pion_theorem}, and $-(16^{\circ}\sim 13^{\circ})\pm6^{\circ}$ based on the quark-line rule~\cite{quark-line-rule}.

\begin{table}[!h]
  \centering
  \caption{The various ratios between ${\boldmath{\psip}} $ and ${\boldmath{\jpsi}}$ decays. $\br_{p \pb \etap}$ and $\br_{p \pb \eta}$ are the branching fractions, $\mf_{p \pb \etap}$ and $\mf_{p \pb \eta}$ are the three-body phase space factors~\cite{pdg_2014}, $\mm_{p \pb \etap}$ and $\mm_{p \pb \eta}$ are the accordingly determined matrix elements. The $|\mm_{p \pb \etap}/\mm_{p \pb \eta}|$ is calculated by $\sqrt{\frac{\br_{p \pb \etap}}{\br_{p \pb \eta}} \cdot \frac{\mf_{p \pb \eta}}{\mf_{p \pb \etap}}}$.}
  \begin{tabular}{lcc}
    \hline
    \hline
     Ratio &$\psip$ &$\jpsi$ \\
    \hline
		$\br_{p \pb \etap}/\br_{p \pb \eta}$ (\%) &$18.3\pm2.5$   &$6.3\pm0.6$ \\
    $\mf_{p \pb \etap}/\mf_{p \pb \eta}$      &0.5            &0.2 \\
    $|\mm_{p \pb \etap}/\mm_{p \pb \eta}|$    &$0.61\pm0.04$  &$0.56\pm0.03$ \\
    \hline
    \hline
  \end{tabular}
  \label{tab-phsp-ratio1}
\end{table}

Table~\ref{tab-phsp-ratio1} shows the details for the ratios of branching fractions, three-body phase space factors, and determined matrix elements. The ratios of the matrix elements are consistent with the theoretical prediction that falls within the range $[0.5,~0.9]$ according to Ref.~\cite{soft_pion_theorem}.

\par Our results for the branching fractions of $\psip\to\p\pb\etap$ and $\jpsi\to\p\pb\etap$ result in the ratio $\frac{\br(\psip\to\p\pb\etap)}{\br(\jpsi\to\p\pb\etap)}=(8.7 \pm 1.0)\%$, where the common uncertainties have been canceled. Even though the ratio is in reasonabe agreement with 12\%, we note that the kinematics of the two processes are very different, and the ``12\%
rule'' may be too naive in this case. The phase space ratio is $\mf_{\psip\to\p\pb\etap}/\mf_{\jpsi\to\p\pb\etap}=8.13$, if any possible intermediate structure is ignored.  Furthermore, the Dalitz plots of $\jpsi$ and $\psip$ decays, shown in Fig.~\ref{fig-dalitz}, indicate that many events in $\psip$ decays, possibly via $N^{\ast} \bar{N} + c.c.$ intermediate states with $p \etap$ or $\pb \etap$ mass greater than 2.13~GeV/$c^2$, are not kinematically possible in $\jpsi$ decays. Taken these factors into account, the $Q$ value is suppressed a lot, implying that the ``12\% rule'' is violated significantly.

\section{ACKNOWLEDGEMENTS}

\par The BESIII collaboration thanks the staff of BEPCII and the IHEP computing center for their strong support. This work is supported in part by National Key Basic Research Program of China under Contract No. 2015CB856700; National Natural Science Foundation of China (NSFC) under Contracts Nos. 11335008, 11425524, 11625523, 11635010, 11735014; the Chinese Academy of Sciences (CAS) Large-Scale Scientific Facility Program; the CAS Center for Excellence in Particle Physics (CCEPP); Joint Large-Scale Scientific Facility Funds of the NSFC and CAS under Contracts Nos. U1532257, U1532258, U1732263; CAS Key Research Program of Frontier Sciences under Contracts Nos. QYZDJ-SSW-SLH003, QYZDJ-SSW-SLH040; 100 Talents Program of CAS; INPAC and Shanghai Key Laboratory for Particle Physics and Cosmology; German Research Foundation DFG under Contracts Nos. Collaborative Research Center CRC 1044, FOR 2359; Istituto Nazionale di Fisica Nucleare, Italy; Koninklijke Nederlandse Akademie van Wetenschappen (KNAW) under Contract No. 530-4CDP03; Ministry of Development of Turkey under Contract No. DPT2006K-120470; National Science and Technology fund; The Swedish Research Council; U. S. Department of Energy under Contracts Nos. DE-FG02-05ER41374, DE-SC-0010118, DE-SC-0010504, DE-SC-0012069; University of Groningen (RuG) and the Helmholtzzentrum fuer Schwerionenforschung GmbH (GSI), Darmstadt.

\end{document}